\documentclass[ba,preprint]{imsart}
\pubyear{TBA}
\volume{TBA}
\issue{TBA}
\firstpage{1}
\lastpage{1}

\usepackage{amsthm}
\usepackage{amsmath}
\usepackage{natbib}
\usepackage{amsmath,amsfonts,amssymb}
\usepackage{booktabs}
\usepackage{comment}
\usepackage[colorlinks,citecolor=blue,urlcolor=blue,filecolor=blue,backref=page]{hyperref}
\usepackage{graphicx}
\usepackage{bm}
\usepackage{subcaption}
\usepackage{placeins}
\usepackage{pdfpages}

\newcommand{\p}{$p_2$\;}
\newcommand{\boldvar}[2][1]{\bm{#2}_{#1}}
\newcommand{\real}[1]{\mathbb{R}^{#1}}

\startlocaldefs
\endlocaldefs

\begin{document}

\begin{frontmatter}
\title{The Multiplex $p_2$ Model: Mixed-Effects Modeling for Multiplex Social Networks}
\runtitle{The Multiplex $p_2$ Model}

\begin{aug}
\author{\fnms{Anni} \snm{Hong}\thanksref{addr1}\ead[label=e1]{annihong@andrew.cmu.edu}},
\author{\fnms{Nynke M.D.} \snm{Niezink}\thanksref{addr1}\ead[label=e2]{nniezink@andrew.cmu.edu}}

\runauthor{Hong \& Niezink}

\address[addr1]{Department of Statistics \& Data Science, Carnegie Mellon University, Pittsburgh, PA 15213, USA \\
    \printead{e1} 
    \printead{e2}
}

\end{aug}
\begin{abstract}
Social actors are often embedded in multiple social networks, and there is a growing interest in studying social systems from a multiplex network perspective. In this paper, we propose a mixed-effects model for cross-sectional multiplex network data that assumes dyads to be conditionally independent. Building on the uniplex \p model, we incorporate dependencies between different network layers via cross-layer dyadic effects and actor random effects. These cross-layer effects model the tendencies for ties between two actors and the ties to and from the same actor to be dependent across different relational dimensions. The model can also study the effect of actor and dyad covariates. As simulation-based goodness-of-fit analyses are common practice in applied network studies, we here propose goodness-of-fit measures for multiplex network analyses. We evaluate our choice of priors and the computational faithfulness and inferential properties of the proposed method through simulation. We illustrate the utility of the multiplex \p model in a replication study of a toxic chemical policy network. An original study that reflects on gossip as perceived by gossip senders and gossip targets, and their differences in perspectives, based on data from 34 Hungarian elementary school classes, highlights the applicability of the proposed method. 

\end{abstract}



\end{frontmatter}





\section{Introduction}

The study of the social behavior of individuals and groups is inseparable from the study of social networks. Social networks can be described by graphs, consisting of nodes, representing social actors (e.g., individuals, countries, organizations), and edges, representing the connections among the actors (e.g., friendship, trade, collaboration). Social actors are often embedded in multiple social networks simultaneously, and the field of social network analysis has long recognized the importance of this multiplexity \citep{borgatti2009network, kadushin2012understanding}. For example, the development of bullying in a classroom cannot be understood without the context of friendships in that classroom \citep[e.g.,][]{rambaran2020bullying} -- the dynamics of positive and negative relations are dependent. Yet, in much applied work, the multiplex network perspective is not leveraged, and single networks are considered in isolation. 

Multiplex networks are structures representing multiple relationships observed among the same group of actors. These relationships can represent heterogeneous edge types, e.g., retweet, follows, and mentions among Twitter users \citep{greene2013producing}, or co-authorship, co-citation, and co-venue relationships among academics \citep{hmimida2015community}. Multiplex networks also play a major role when studying network perspectives. Self-reported social networks might be imprecise measurements of underlying social behavior, and social actors can have different perspectives on a relationship (e.g., $i$ may experience behavior by $j$ as bullying but $j$ does not think they bully $i$). Social desirability bias too can drive individuals to adjust their reports (e.g., $i$ reports less gossip than $i$ actually participated in). \cite{Tatum2020} studied the difference in perspectives between bullies and victims in school classes by aggregating the two networks with different perspectives into a single disagreement network. While this approach allows us to study discrepancies, it does not provide an understanding of the context in which they arose. A multiplex network approach would be a more natural way to study the multiple perspectives (relationships) as well as their discrepancies.

In this paper, we develop a multiplex network model that would allow for such an analysis: a conditional independent dyads mixed-effects model for cross-sectional, directed binary network data. Network dyads are given by pairs of nodes and the configuration of relations among them. Conditional dyad independence is a popular assumption to enable modeling complex network data and gives rise to a rich class of models, such as latent space models \citep{hoff2002latent} and stochastic blocks models \citep{holland1983stochastic}. We develop a model in the tradition of the  \p  modeling framework \citep{Duijn2004}, which represents the probability of a dyad outcome in terms of baseline network density, reciprocity, and actor and dyad heterogeneity. While its predecessor, the $p_1$ model \citep{holland1981exponential}, models actor heterogeneity with fixed effects representing actors' differential tendencies to send and receive ties (thus giving rise to a large number of parameters for large networks), the \p model represents these tendencies by random effects and incorporates covariates to model actor and dyadic heterogeneity. For its interpretability and simplicity, the \p model has been widely used, e.g., to study physician communication patterns  \citep{keating_factors_2007} and advice seeking between public schools \citep{spillane2015intra}. The multilevel  extension of the \p model \citep{zijlstra_multilevel_2006} enabled a large number of multi-group network studies \citep[e.g.,][]{Veenstra2007, vermeij2009ethnic, Tolsma2013, smith2014ethnic, zijlstra2011multilevel}.

Here, we extend  the \p model for multiplex network analysis. The parameters and characteristics of the uniplex \p model remain as within-network effects in the multiplex model. Yet, the multiplex model introduces new cross-network density and cross-network reciprocity parameters, modeling the tendencies for ties between two actors to be dependent across different relational dimensions. Actors' random sender and receiver effects for the different relational dimensions are modeled as different, but dependent.

\cite{bellio2021maximum} proposed a maximum likelihood-based approach to \p model estimation. However, a robust Bayesian estimation procedure is still lacking. The \texttt{R} package \texttt{dyads} \citep{dyads} is the only existing Bayesian implementation of the \p model, but has a few limitations. These include its inability to include multiple actor covariates, its lack of convergence checks and goodness-of-fit tests, and its inability to handle missing network data. Accompanying this paper, we developed a Bayesian implementation of the multiplex \p model in the \texttt{Stan} probabilistic programming framework \citep{stan2022}. This method includes the standard (uniplex) \p model as a special case.  Our \texttt{R} package \texttt{multiP2} includes multiplex goodness-of-fit measures and the implementation can handle missing network data. 

In the following, we will first give a brief overview of existing methods for cross-sectional multiplex network data (Section \ref{ss:literature}) and then present the multiplex \p model (Section \ref{sec:multiplex}) and its estimation (Section \ref{sec:estimation}). We then introduce goodness-of-fit measures for multiplex networks, which can be used in posterior predictive checks (Section \ref{sec:gof}). Section \ref{sec:sim} presents several simulation studies in which we evaluate our choice of priors, the accuracy of our estimation procedure, and the inferential properties of the method. Sections \ref{sec:pol} and \ref{sec:gossip} present two applications. The first one replicates the analysis by \cite{leifeld2012information} of how information is exchanged in policy networks, and illustrates that our multiplex approach offers additional insights into reciprocation patterns across different types of information (political and  scientific). The second study analyzes the discrepancies in reports on gossip ties from the viewpoint of the gossiper and the gossip target based on data not previously studied. We find that students are not particularly perceptive when it comes to identifying individuals who gossip about them. However, they tend to gossip about those whom they believe are gossiping about them. We conclude with a discussion in Section \ref{sec:disc}.

\subsection{Multiplex network methods} \label{ss:literature}

Several methods have been developed for the analysis of cross-sectional multiplex network data. Network regression techniques such as the multiple
regression quadratic assignment
procedure \citep[MRQAP;][]{krackhardt1988predicting, dekker2007sensitivity} allow us to model a network as a function of other networks (c.f.\ multiple regression) but do not simultaneously model multiple networks as the outcome (c.f.\ multivariate regression). The multiplex stochastic block model \citep{barbillon2017stochastic} detects communities with information from multiple network layers, and a recent development allows the number of communities to vary in different layers \citep{amini2024hierarchical}. These methods are tailored to assign group membership to social actors based on network data, but do not take into account the role of covariate information. The social relations model (SRM) does take this into account and supports the simultaneous analysis of multiple continuous-valued networks \citep{nestler2018likelihood} but not that of binary networks. Multiplex latent space models \citep{salter2017latent, sosa2022latent} can be used to model continuous or binary networks, but differ from the \p model  in terms of how the probability of a tie is modeled. The \p model explicitly models network characteristics such as density and reciprocity while the latent space model assumes that unobserved actor attributes, as represented by latent positions, affect the connectivity pattern. Recent developments in exponential family random graph models (ERGMs) allow for the modeling of two or more binary networks as outcomes simultaneously \citep{wang2013exponential, wang2016social,chen_statistical_2021}. Yet, ERGM estimation can be computationally expensive and model convergence is hard to achieve in practice with increasing network dimensions. The stochastic actor-oriented model (SAOM) was developed to analyze the dynamics of network data \citep{Snijders2001}, and extended for the coevolution of multiple binary networks \citep{snijders2013model}. Although it is possible to study cross-sectional network data using a stationary SAOM, the applicability and properties of stationary SAOMs, even for uniplex network data, remain unexplored.

\section{The multiplex $p_2$ model}
\label{sec:multiplex}
Suppose there are $T$ layers of networks on the same set of $n$ actors. Let $\bm{M} \in \{0,1\}^{n \times n \times T}$ denote the directed binary multiplex network among these actors. We assume that this network has no self-loops (i.e., $i$ can not send ties to themselves; $M_{iit} = 0$ for all $i$ and $t$). We denote the $p$ dyadic covariates (e.g., absolute age difference) by $\bm{Z} \in \real{n \times n \times p}$ and the $q$ actor covariates  (e.g., gender) by $\bm{X} \in \real{n \times q}$. In the following, we treat $p$ and $q$ as the same for all effects, without loss of generality. 
If we let $m^{t}_{ij} = 1$ denote that there is an edge from $i$ to $j$ in layer $t$, and $m^{t}_{ij} = 0$ if there is no such edge, we can represent the (multiplex) outcome on dyad $\{i,j\}$ by $M_{\{ij\}}  = \{m^{1}_{ij}, m^{1}_{ji}, \dots, m^{T}_{ij}, m^{T}_{ji}\}$. Since the \p model is a conditionally independent dyads model, we can decompose the probability of a network $\bm{M}$ into the probabilities of the dyad outcomes $M_{\{ij\}}$.  

Within each layer, the multiplex \p model accounts for the overall propensity of ties by a density parameter, and for the likelihood of a tie being reciprocated between two actors by a reciprocity parameter. We denote the the density and reciprocity parameters for layer $t$ as $\mu^t$ and $ \rho^t$, and let $\bm{\mu} = \{\mu^1, \dots, \mu^T\}$ and $ \bm{\rho} = \{\rho^1, \dots, \rho^T\}$ .
The model also allows us to evaluate the effects of network covariates on density and reciprocity. If we denote the corresponding coefficients by $\bm{\delta}_{\mu}^t\in \real{p}$ and $\bm{\delta}_{\rho}^t \in \real{p}$, the total density and reciprocity effects for dyad $\{i,j\}$ in layer $t$ are
\begin{equation}
\begin{split}
\mu_{ij}^t &= \mu^t + (\bm{Z}_{ij})^{\top}\bm{\delta}_{\mu}^t,\\
\rho_{ij}^t &= \rho^t + (\bm{Z}_{ij})^{\top}\bm{\delta}_{\rho}^t.
\end{split}
\end{equation}
Across layers, we can model the tendency for one actor to establish multiple connections to another actor (e.g., $i$ considers $j$ a friend \emph{and} a collaborator) by a cross-layer density effect. We can also capture cross-layer reciprocity -- the tendency for actors to reciprocate an incoming tie of one kind (e.g., political information) with an outgoing tie of another kind (e.g., scientific information). We let $\mu_{\text{cross}}^{(t,s)}$ denote the baseline cross-layer density effect for layers $s$ and $t$ and $\rho_{\text{cross}}^{(t,s)}$ the baseline cross-layer reciprocity, with $\bm{\mu}_\text{cross}$ and $\bm{\rho}_\text{cross}$ containing the cross-layer density and cross-layer reciprocity parameters for all pairs of layers. The cross-network density and reciprocity too can depend on dyadic covariates, with corresponding coefficients $\bm{\delta}_{\mu_\text{cross}}^{(t,s)} \in \real{p}$ and $ \bm{\delta}_{\rho_\text{cross}}^{(t,s)} \in \real{p}$. In the model, the total cross-layer density and reciprocity effects for dyad $\{i,j\}$ in layers $s$ and $t$ are thus given by
\begin{equation}
\label{eq: cross-eff}
\begin{split}
    \mu_{\text{cross},ij}^{(t,s)} &= \mu^{(t,s)}_\text{cross} + (\bm{Z}_{ij})^\top\bm{\delta}_{\mu_\text{cross}}^{(t,s)},\\
    \rho_{\text{cross},ij}^{(t,s)} &= \rho^{(t,s)}_\text{cross} + (\bm{Z}_{ij})^\top \bm{\delta}_{\rho_\text{cross}}^{(t,s)}.
\end{split}
\end{equation}

The \p model captures actors' differential tendency to send ties by random sender effects, and their differential tendency to receive ties by random receiver effects. The tendencies for actors to send and receive ties can depend on individual actor characteristics. We define different actor sender and receiver effects for each layer, and allow actors' tendencies to send and to receive ties to be correlated across relationships. For example, if a person sends texts to a lot of people, they may also send emails to a lot of people. Yet, we assume that one actor's tendency to send or receive ties does not affect this tendency of the other actors  -- the random effects are correlated within but not between individuals. Formally, we let $\alpha_i^t $ and $\beta_i^t $ denote the tendencies for actor $i$ to send and receive ties in network layer $t$, respectively. The sender and receiver effects for all actors in layer $t$ are $\bm{\alpha}^t = \{\alpha_1^t, \dots, \alpha_n^t\}$ and $ \bm{\beta}^t = \{\beta_1^t, \dots, \beta_n^t\}$. The coefficients $\bm{\gamma}_{\alpha}^t \in \mathbb{R}^{q}$ and $\bm{\gamma}_{\beta}^t \in \mathbb{R}^{q}$ represent the effects of the individual-specific covariates $\bm{X}$ on actors' tendencies to send and receive ties. Let $\bm{C}_i = [A^1_i, B^1_i, ..., A^T_i, B^T_i]^\top, \bm{C}_i \in \mathbb{R}^{2T}$ denote the collection of random actor effects for actor $i$ in $T$ layers of networks. We assume that the random effects follow a multivariate Gaussian distribution such that $\bm{C}_i \sim \mathcal{N}(\bm{0}, \bm{\Sigma_{AB}}).$ With $\bm{A}^t = [A^t_i, ..., A^t_n]^\top, \bm{B}^t = [B^t_i, ..., B^t_n]^\top$, the actor sender and receiver effects are  
\begin{equation}
\label{eq: multi actor}
\begin{split}
        \bm{\alpha}^t = \bm{X}\bm{\gamma}_{\alpha}^t + \bm{A}^t,\\
    \bm{\beta}^t = \bm{X}\bm{\gamma}_{\beta}^t + \bm{B}^t.
\end{split}
\end{equation}
As the actor effects are assumed to be independent across individuals, we have that $\text{cov}(A^t_i, A^s_j) = \text{cov}(B^t_i, B^s_j) = \text{cov}(A^t_i, B^s_j) = 0$, for all dyads $\{i, j\}$ and layers $t$ and $s$. 

Now we can bring all the parts together and define the probability function on a dyad in a $T$-plex network. If we define
\begin{equation}
\label{eq: multi prob numer}
    \begin{aligned}
        K_{ij}(M_{\{i,j\}})  &= 
\sum_{t = 1}^T \Big( {m^{t}_{ij}(\mu_{ij}^t + \alpha_i^t + \beta_j^t) + m^{t}_{ji}(\mu_{ji}^t + \alpha_j^t + \beta_i^t) + m^{t}_{ij} m^{t}_{ji} \rho_{ij}^t} \Big)  \\
&\phantom{{}={}} + \sum_{t,s = 1,...,T \,:\, t < s }{(m_{ij}^{t}m_{ij}^{s}}  + m_{ji}^{t}m_{ji}^{s})\,\mu_{\text{cross},ij}^{(t,s)}  \\
&\phantom{{}={}} + \sum_{t,s =1,...,T \,:\, t < s }{(m_{ij}^{t}m_{ji}^{s}  + m_{ji}^{t}m_{ij}^{s})\,\rho_{\text{cross},ij}^{(t,s)}}\,,
\end{aligned}
\end{equation}
the probability of outcome $M_{\{i,j\}}$ on dyad $\{i,j\}$ is given by
\begin{equation}
\label{eq: multi prob}
    P(M_{\{i,j\}}) = \frac{\exp\{K_{ij}(M_{\{i,j\}})\}}{\sum_{G_{\{i,j\}} \in \{0,1\}^{2T}} \exp\{K_{ij}(G_{\{i,j\}})\}},
\end{equation}
where the denominator sums over all possible realizations $G_{\{i,j\}}$ of the $T$ relationships on this dyad. 

\section{Model estimation}
\label{sec:estimation}
The iterated generalized least-squares procedure \citep{Duijn2004} was one of the earliest proposed methods for parameter estimation in the \p model. Subsequently, \cite{zijlstra2011mcmc} explored several Markov chain Monte Carlo estimation procedures for the \p model, two of which used random walk proposals. More recently, \cite{bellio2021maximum} proposed a maximum likelihood estimation procedure based on the Laplace approximation.

In this study, we employ Hamiltonian Markov chain Monte Carlo (HMCMC) \citep{duane1987hybrid, neal2011mcmc, betancourt2011geometry} for estimating the parameters of the multiplex \p model. HMCMC significantly improved the sampling performance compared to random walk Metropolis in situations where hierarchical structures, such as random-effect models, induce correlations between global and local parameters \citep{betancourt2015hamiltonian}. The advantage of HMCMC lies in its ability to effectively explore the parameter space by leveraging the local curvature of the target distribution. However, HMCMC requires careful fine-tuning, including specifying the derivative of the target distribution. To address this, we utilize the robust implementation of HMCMC provided by the \texttt{Stan} probabilistic programming language \citep{carpenter2017stan, stan2022}. By leveraging the HMCMC algorithm in \texttt{Stan} through the \texttt{R} package \texttt{rstan} \citep{rstan}, we can effectively handle the  structure of the multiplex \p model.

Next, we present the likelihood function and the prior distributions for the multiplex \p model. There are $2^{2T}$ possible outcomes on a dyad in a $T$-plex network. Dyadic outcome $M_{\{i,j\}}$ follows a categorical logit distribution, 
\begin{equation}
\begin{split}
        M_{\{i,j\}}&\, |\, \bm{\mu}, \bm{\rho}, \boldvar[\text{cross}]{\mu},\boldvar[\text{cross}]{\rho}, \\
    &\bm{Z}, \bm{\delta}_{\mu},\bm{\delta}_{\rho}, \bm{\delta}_{\mu_{\text{cross}}},\bm{\delta}_{\rho_{\text{cross}}},\\
    &\bm{X}, \bm{\gamma}_{\alpha}, \bm{\gamma}_{\beta} ,\bm{C}\\
    &\sim \text{Categorical} (M_{\{i,j\}}|P(M_{\{i,j\}})), 
\end{split}
\end{equation}
where $P(M_{\{i,j\}})$ is as defined in Equation \eqref{eq: multi prob}. We assume normal priors for the baseline within-network parameters $\bm{\mu}^t$ and  $\bm{\rho}^t$ and cross-network parameters $\boldvar[cross]{\mu}^{(t,s)}$ and $\boldvar[cross]{\rho}^{(t,s)}$,
\begin{equation}
        \bm{\mu}^t, \bm{\rho}^t, \boldvar[cross]{\mu}^{(t,s)}, \boldvar[cross]{\rho}^{(t,s)} \sim \mathcal{N}(0, 10),
\end{equation}
and set the priors for the fixed network and actor effects of the covariates to
\begin{equation}
      {\theta}_{k} \sim \mathcal{N}\left(0, \frac{10}{\sigma(\bm{g}_k)}\right),
\end{equation}
where ${\theta}_k \in \{\bm{\delta}_{\mu}^\top,\bm{\delta}_{\rho}^\top, \bm{\delta}_{\mu_\text{cross}}^\top, \bm{\delta}_{\rho_\text{cross}}^\top, \bm{\gamma}_{\alpha}^\top, \bm{\gamma}_{\beta}^\top\}$ and $\bm{g}_k$ is the corresponding covariate and $\sigma(\bm{g}_k)$ the sample standarded deviation of the covariate. As the absolute values of $\bm{\mu}$, $\bm{\rho}$, $\boldvar[cross]{\mu}$, $\boldvar[cross]{\rho}$, and $\bm{\theta}_k\bm{g}_k$ are unlikely to exceed 10 \citep{zijlstra2011mcmc}, we pick weakly informative priors with mean 0 and standard deviation 10.

In the multiplex \p model, the random effect $\bm{C}_i \in \real{2T}$ follows a multivariate Gaussian distribution. There are several potential choices for the prior of the variance-covariance matrix $\bm{\Sigma_{AB}}$. It is common to assume an inverse Wishart distribution, as was done by \cite{zijlstra2011mcmc} for the univariate \p model, because of its conjugate property. However, many issues of the inverse Wishart distribution have been pointed out, most notably, that the variance and the correlation parameters are correlated \citep[e.g.,][]{akinc2018bayesian, tokuda2011visualizing, liu2016comparison}. Since \texttt{Stan} does not require the use of conjugate priors, we will use the strategy first proposed by \cite{barnard2000modeling} and studied by \cite{akinc2018bayesian} and \cite{tokuda2011visualizing}.
That is, we first decompose the covariance matrix $\bm{\Sigma_{AB}}$ into a vector of coefficient scales $\bm{\sigma}$ and a correlation matrix $\bm{\Omega}$, 
\begin{equation}
    \bm{\Sigma_{AB}}= \text{diag}(\bm{\sigma}) \times \bm{\Omega} \times \text{diag}(\bm{\sigma}), 
\end{equation}
%
and let the scale and correlation components have different, independent priors. We set the prior of the elements of the scale vector to an inverse gamma distribution, and let the correlation matrix follow an LKJ correlation distribution \citep{lewandowski2009generating} with shape parameter $\eta = 2$,
\begin{equation}
\begin{split}
    &\sigma_i \sim \text{InverseGamma}(\alpha = 3, \beta = 50) \\
    &\bm{\Omega} \sim \text{LJKCorrelation}(\eta = 2).
\end{split}
\end{equation}
In the LKJ distribution, $\eta = 1$ generates a uniform correlation distribution. For $\eta > 1$, the distribution puts more mass on the unit matrix, representing less correlation. For $\eta < 1$, mass concentrates away from the unit matrix thus favoring matrices with more correlation. We let $\eta$ be slightly larger than 1 (i.e., $\eta = 2$), because most of the weight should be on the unit matrix but some correlation among actors' roles across networks in terms of sending and receiving ties is expected. We picked the hyperparameters $\alpha = 3$ (shape) and $\beta = 50$ (scale) for the priors of $\sigma_i$ based on the results of the prior predictive checks in Section \ref{sec:sim}.  

\section{Goodnesss-of-fit on multiplex networks}
\label{sec:gof}

A useful model should capture the key structure of the data-generating process such that the fitted model generates data similar to the observed data. In this section, we discuss existing methods for assessing uniplex network model fit and propose measures for multiplex goodness-of-fit analysis. We illustrate these methods in the application in Section \ref{sec:pol}.  

\paragraph{Uniplex goodness-of-fit.} Simulation-based goodness-of-fit methods are widely used in social network research to evaluate if a model has captured the characteristics of the observed networks \citep{hunter2008goodness}, and have been adapted to the Bayesian framework in the form of posterior predictive checks \citep{caimo2011bayesian}. In the latter, the observed network data are compared to a set of networks simulated from the posteriors of the parameters, on key network statistics such as the indegree and outdegree distribution, and the triad census \citep{davis1967structure, holland1977method}. The indegree and outdegree distribution are the distribution of actors' numbers of incoming and outgoing ties, respectively. Their goodness-of-fit analyses help assess whether the model captures the tendencies of actors in a social network to send and receive ties. 
The triad census counts the frequency of each of the sixteen possible network configurations among three actors present in a directed network. Goodness-of-fit analysis on the triad census helps assess the extent to which triadic effects (e.g., transitivity) and local group structure are accurately represented by the model. 
Other potential goodness of fit measures include network density (the proportion of potential ties that are actually present), reciprocity (the proportion of ties that are reciprocated) the distribution of geodesic distances (lengths of shortest paths between actors in a network) and measures of network autocorrelation \citep[e.g.,][]{moran1948interpretation,geary1954contiguity}. Generally, it is good practice to align network goodness-of-fit measures with the applied research question studied. For example, if a researcher is interested in the effect of performance homophily on friendship network formation, it makes sense to evaluate a model's goodness-of-fit on network autocorrelation on performance.

\paragraph{Multiplex goodness-of-fit.} In line with this rationale, we propose statistics to describe the fit of multiplex network models. Although we illustrate these only in the context of the multiplex \p model, they are applicable for multiplex network models generally. Naturally, for each layer of the observed multiplex network, we can evaluate goodness-of-fit on the statistics described above for uniplex networks. 
Moreover, for every pair of layers, we can measure the tendency for ties to occur jointly and to be reciprocated between layers. We can measure the tendency for ties to occur in two layers simultaneously by the Jaccard index of the two layers -- the proportion of ties that appear in both networks over the total number of unique ties in the two networks. If we let ${\bm M}^t \in \{0,1\}^{n\times n}$ denote the adjacency matrix corresponding to layer $t$ of multiplex network ${\bm M}$, the Jaccard index between layers $s$ and $t$ is given by %
\begin{equation}\text{Jaccard}({\bm M}^t, {\bm M}^s)= \frac{|X_t \cap X_s|}{|X_t/X_s| + |X_t/X_s| + |X_t \cap X_s|},
\label{eq:Jaccard}
\end{equation}
where $X_t$ and $ X_s$ denote the sets of edges in layers $t$ and $s$, respectively. We can evaluate cross-layer reciprocity by Jaccard(${\bm M}^t, ({\bm M}^s)^\top$). To capture the (dis)similarities of actors' tendencies to send and receive ties in different network layers, we can use the correlations among actors' indegrees and outdegrees in all layers. Note that each of these statistics aligns with an effect in the multiplex \p model.  

\section{Simulation studies}
\label{sec:sim}
In this section, we present the results of several simulation studies assessing the appropriateness of the selected priors for social networks, the accuracy of the algorithm used for computation, and the inferential properties of the proposed method, following the workflow discussed by \cite{gelman2020bayesian} and \cite{schad2021toward}. To this end, we simulate networks with a biplex structure ($T = 2$) consisting of 30 actors and thus ${ 30 \choose 2} = 435$ dyads. Networks of this size are typical in educational applications. The policy-related network in Section \ref{sec:pol} is of this size as well. The data are generated outside our estimation environment.  

\paragraph{Prior predictive checks.} To determine whether a model with a given set of priors on the parameters is consistent with domain expertise, we conduct prior predictive checks, examining if the simulated data generated from the priors is representative of the expected data. We analyze three key social network statistics: density, reciprocity, and transitivity. Like density and reciprocity (described in Section \ref{sec:gof}) transitivity is another key network statistics, which refers to the proportion of nodes with a common connection that are also connected. Reciprocity and transitivity can capture the human tendencies to reciprocate positive ties (e.g., liking) and to operate in small groups  \citep{robins_doing_2015}. All three statistics are bounded between 0 and 1 and have context-dependent expected ranges. For example, friendship networks often have high reciprocity and transitivity, while networks with implicit hierarchies, such as advice-seeking, tend to have low reciprocity. Density tends to decrease with increasing network size. We aim to choose weakly informative priors that incorporate this prior knowledge while allowing the data to drive the inference. 

\begin{figure}
    \centering
    \includegraphics[scale=0.6]{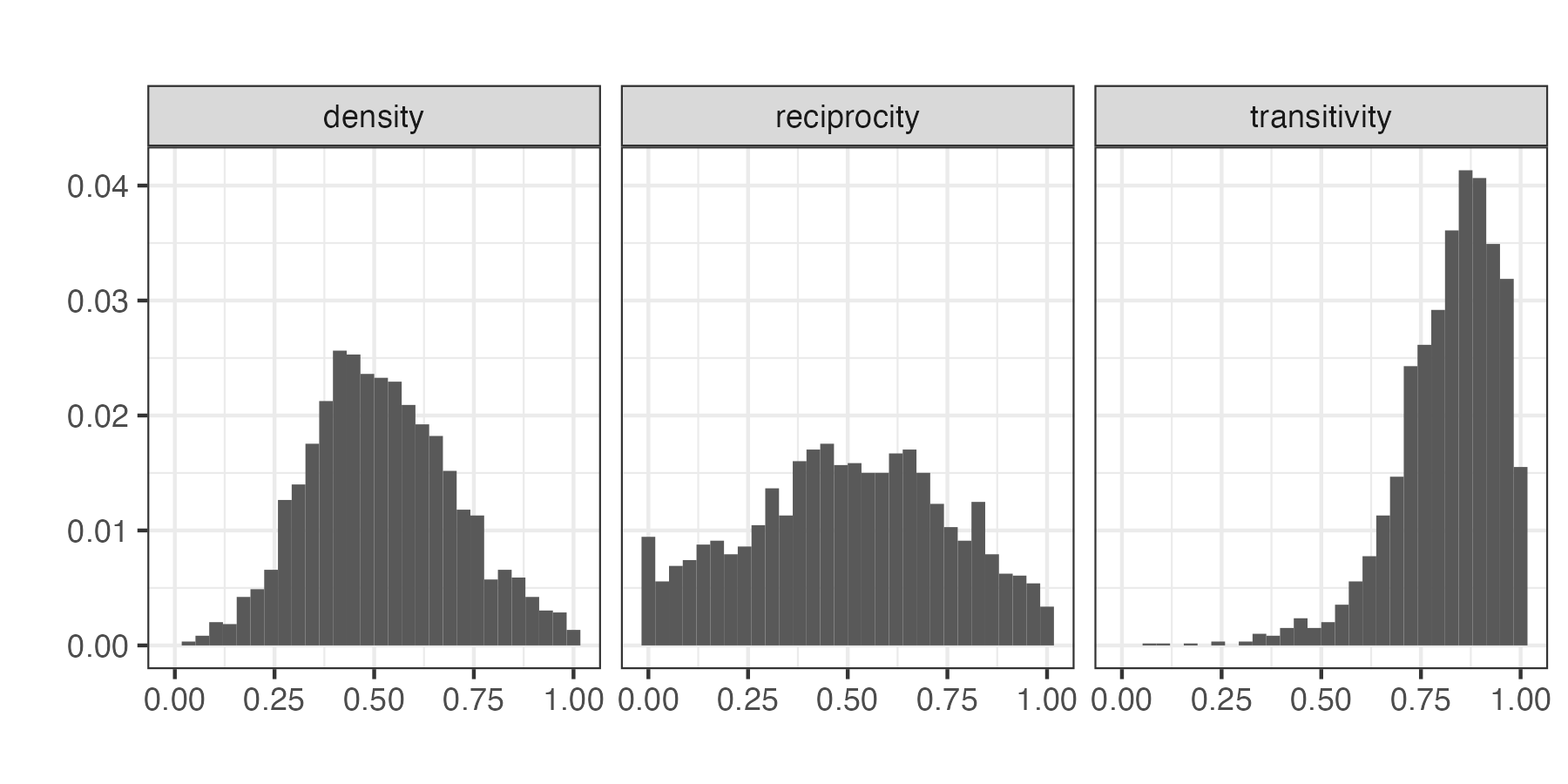}
    \caption{Prior predictive checks: density, reciprocity, and transitivity calculated on 1000 networks simulated based on the priors defined in Section \ref{sec:estimation}.}
    \label{fig:ppc}
\end{figure}

Figure \ref{fig:ppc} shows the prior predictive check results based on 1000 network draws from the prior defined in Section \ref{sec:estimation}. The simulated density and reciprocity cover a wide range of values with peaks at 0.5, coinciding with the zero means of the priors for the density and reciprocity parameters $\mu$ and $\rho$. The simulated transitivity skews higher than typical social networks (0.3 to 0.6 is typical in, e.g., friendship networks), but it still covers a wide range of values. Note also that this statistic is not explicitly modeled by the \p model. Overall, the prior predictive checks yield adequate results, especially when we compare them to the results obtained with the inverse Wishart prior, which used to be the default for the \p model \citep{zijlstra2011mcmc}. For the prior predictive check results using the inverse Wishart prior, see Figure 1 of the Supplementary Material \citep{hong2023supplement}.   

\begin{figure}[h]
    \centering
    \includegraphics[scale=0.67]{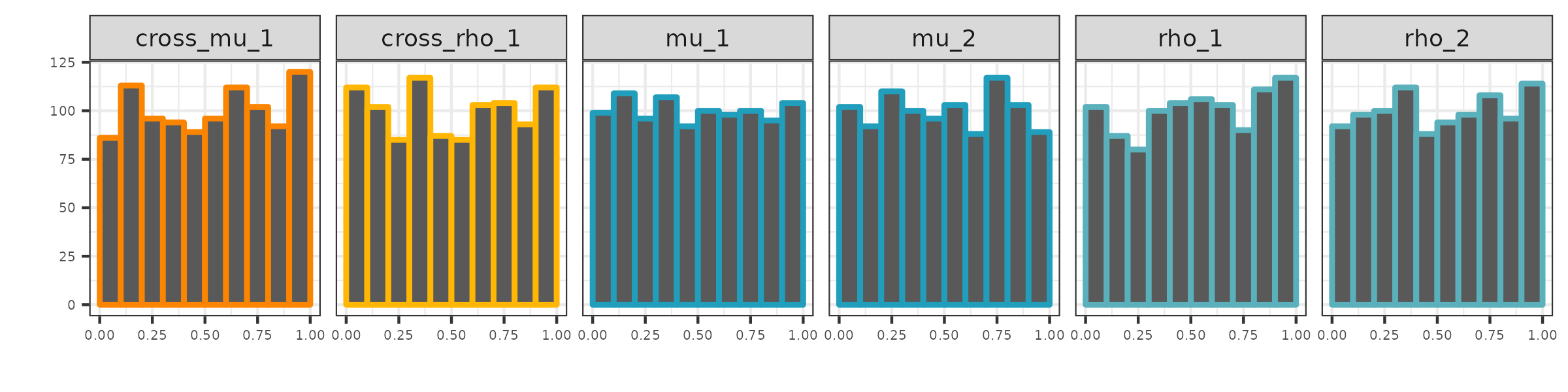}
    \caption{Simulation-based calibration: the distribution of the percentile statistics of the model parameters appear to be uniform, which is consistent with inference being well calibrated.}
    \label{fig:rank}
\end{figure}

\paragraph{Computational faithfulness.}
We conduct simulation-based calibration (SBC) to assess the soundness of our posterior sampler \citep{talts2020validating}. This procedure relies on the fact that the posterior distribution estimated from data generated from the prior should resemble the prior distribution, on average. We first draw $L = 1000$ random samples from the prior distribution, $\tilde{\bm{\theta}}_l \sim \pi(\bm{\theta}), l = {1, ..., L}$, and for each prior draw we generate a biplex network $\tilde{\bm{M}}_l$. We then fit the multiplex \p model to each simulated network $\tilde{\bm{M}}_l$, thus obtaining $L$ estimated posterior distributions. If the model is well-calibrated, the sample from the prior could fall anywhere on the corresponding estimated posterior distribution. In other words, the rank of parameter $\tilde{\theta}_{lj}$ with respect to a given number of $K = 100$ draws from the posterior $\{\theta_{lj,1}^{'}, \dots, \theta_{lj,K}^{'}\}$ should be uniformly distributed.  Figure \ref{fig:rank} shows the histograms of the percentile statistics of the model parameters. These are uniformly distributed, so we conclude that our posterior sampler is sound.

\begin{figure}
    \centering
    \includegraphics[scale=0.6]{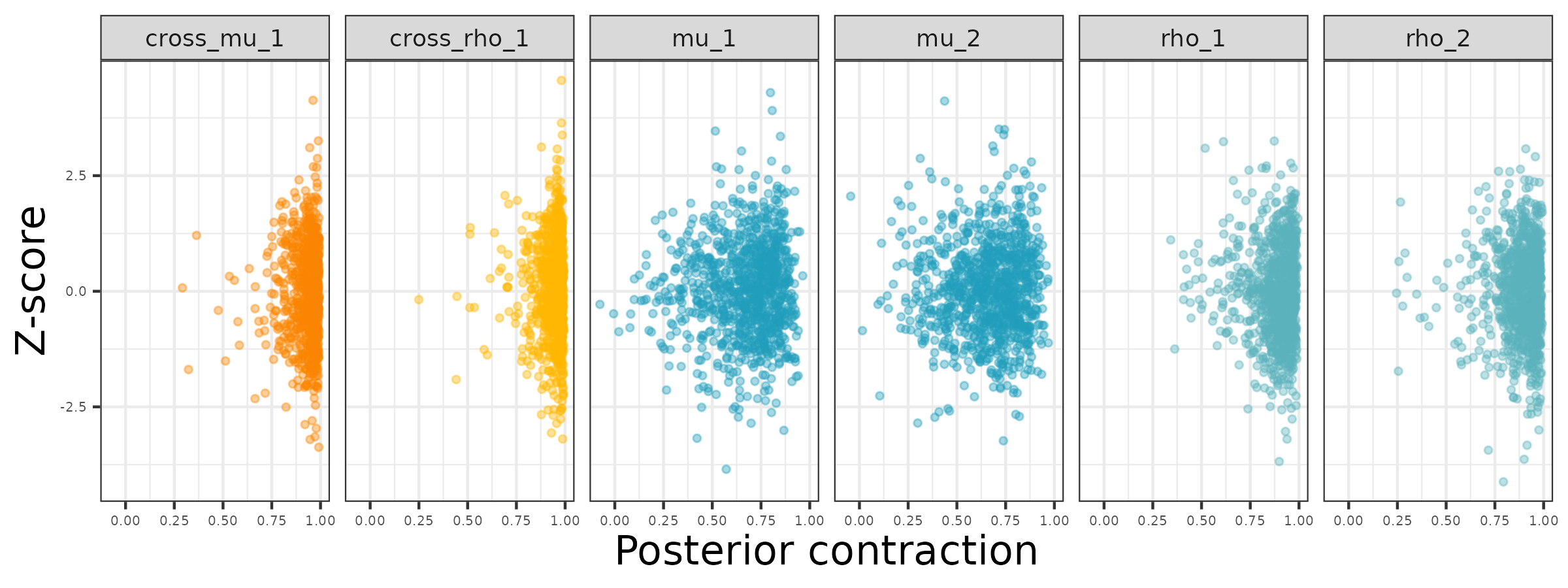}
    \caption{Model sensitivity: plots of z-scores against the posterior contraction calculated on 1000 simulations shows moderately low bias and low variance.}
    \label{fig:modelsensitivity}
\end{figure}

\vspace{-5mm}
\paragraph{Model sensitivity.}
Finally, we assess the model's adequacy for inference by measuring the bias of the posterior mean and the variance reduction from the prior to the posterior. We repeat the simulation steps from the computational faithfulness section and obtain the posterior estimates. For each parameter, the bias can be summarized by a posterior $z$-score: $z = \frac{\mu_\text{post} - \tilde{\theta}}{\sigma_\text{post}},$ where the posterior mean $\mu_\text{post}$ is compared to the true parameter value $\tilde{\theta}$, scaled by the posterior uncertainty. We estimate the reduction in uncertainty by the posterior contraction $s = \frac{\sigma^2_\text{prior} - \sigma^2_\text{post}}{\sigma^2_\text{prior} }$, which is
the variance reduction compared to the prior variance. \cite{schad2021toward} recommend plotting the posterior $z$-score against the posterior contraction for each parameter. Ideally, a model with low bias and low variance would have all the points concentrated around a posterior $z$-score of $0$ and a posterior contraction of 1. Figure \ref{fig:modelsensitivity} shows that the posterior $z$-scores are centered around $0$ with concentration at the the level of posterior contraction close to 1, indicating the multiplex \p model has moderately low bias and low variance. We note that the two density parameters ($\mu_1, \mu_2$) have higher variance than the rest of the parameters.

\section{Application: Information exchange in a policy network}
\label{sec:pol}

In this illustrative study, we analyze the patterns of information exchange and the perception of influence among 30 actors in the policy domain of toxic chemicals regulation in Germany in the 1980s.
This (multiplex) policy network was studied previously by \cite{leifeld2012information}, to determine how governmental and nongovernmental actors chose their potential interaction partners in this historic context, though not from a multiplex network perspective.

\begin{figure}
    \centering
\includegraphics[scale=0.5]{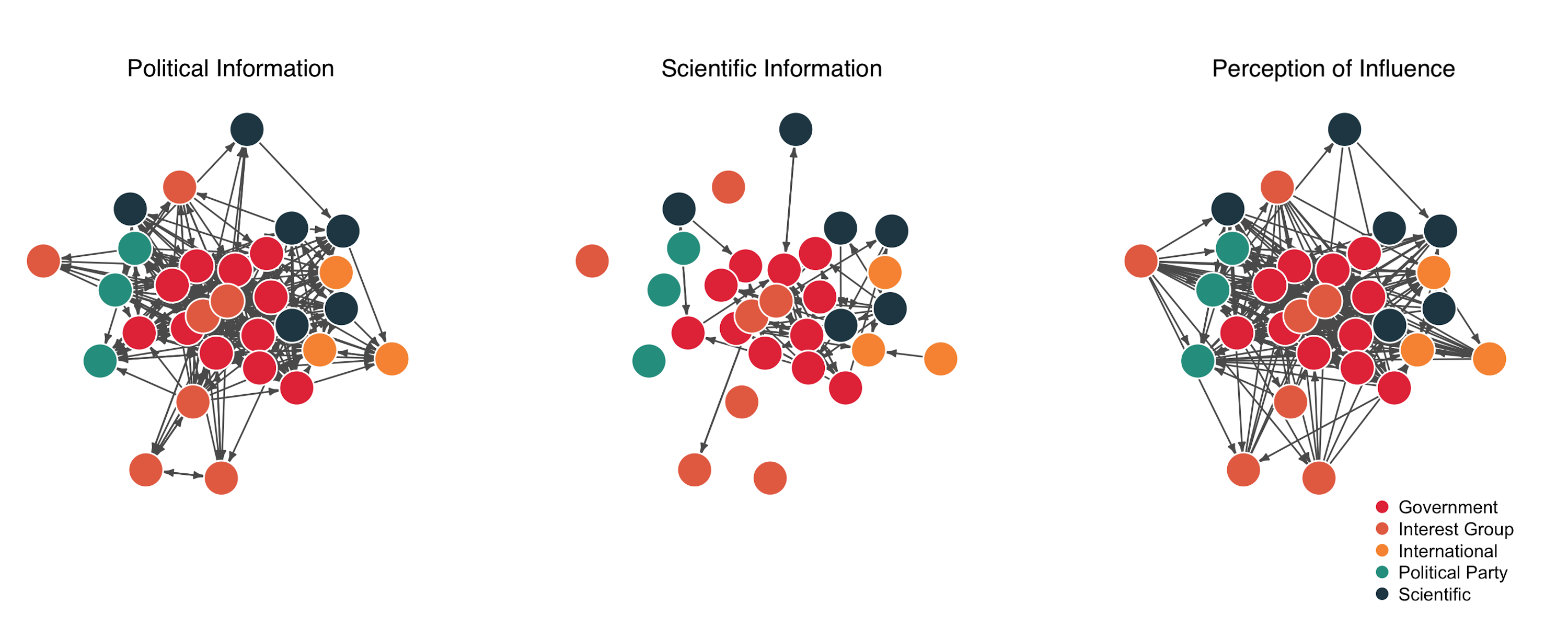}
    \caption{The information exchange and perception of influence networks of organizations in the policy domain of toxic chemicals regulation
in Germany in the 1980s.  Node color indicates institution type. } 
    \label{fig:chem_networks}
\end{figure}

\cite{leifeld2012information} studied two types of information exchange networks and the network of organizations' perception of influence. In the \textit{political information} network, an edge from organization $i$ to organization $j$ indicates that $i$ perceived $j$ as a partner in exchanging political information regarding chemical controls. This network was obtained by asking organizations to list the names of all organizations with whom they regularly exchanged information about affairs related to chemicals control. An edge in the \textit{scientific information} network indicates that organization $i$ provided scientific and technical information about toxic chemicals to organization $j$. In the \textit{perception of influence} network,  an edge from organization $i$ to organization $j$ indicates that $i$ perceived  $j$ to be particularly influential in the policy making process. Figure \ref{fig:chem_networks} depicts the three networks. We refer to \cite{schneider1988politiknetzwerke} and \cite{leifeld2012information} for further details about the selected actors and the data collection process. Given that power dynamics likely impact information exchange and perception of influence, we take into account the effect of governmental status (indicating if an organization is governmental) in our analyses. An organization is also likely to perceive another organization as more influential and engage in greater information exchange if they hold similar policy positions. Therefore, we account for policy preference similarity in this study -- see \cite{leifeld2012information} for the definition of this dyadic covariate.

We address two of the main hypotheses by \cite{leifeld2012information} in this study. Firstly, we assess whether information exchange is more likely among organizations with pre-existing communication ties. We expect ties will co-occur in the two information networks, and to be reciprocated within and across these two network layers (Hypothesis 1). Secondly, we investigate whether institutions are more likely to exchange information with institutions they perceive as influential (Hypothesis 2), as the potential benefits gained from high-influence actors may justify the costs associated with establishing information ties. 

\cite{leifeld2012information} tested these hypotheses by fitting separate (uniplex) exponential random graph models to the political and scientific information networks, including the other two networks as covariates to examine cross-layer effects. By contrast, here we fit a multiplex \p model that simultaneously models the three networks as outcomes. As both directions of cross-layer effects (that is, density \emph{and} reciprocity) are incorporated in our model, this approach naturally extends the idea of using existing communication channels to cross-layer reciprocity, going beyond what was analyzed in the original study. Moreover, our random actor-level effects may yield insights into the association of actors' behaviors within different relational contexts.

\begin{table}[!b] \centering 
\begin{tabular}{lcccccc} 
\toprule
& & & \multicolumn{4}{c}{Jaccard distance to:}\\
Network & Density & Reciprocity & \textsc{sc} & \textsc{sc}$^\top$ &  \textsc{pe} & \textsc{pe}$^\top$\\ 
\midrule
Political information (\textsc{po}) & $0.393$ & $0.614$ & 0.174 & 0.157 & 0.398 & 0.308 \\ 
Scientific information (\textsc{sc}) & $0.072$ & $0.349$ &  &  & 0.123 & 0.078 \\ 
Perception of influence (\textsc{pe}) & $0.325$ & $0.382$ \\ 
\bottomrule
\end{tabular} 
  \caption{Descriptive statistics of the information exchange and perception of influence networks. The Jaccard distance between networks \textsc{x} and \textsc{y} reflects cross-layer network overlap and is determined as in equation \eqref{eq:Jaccard}. The Jaccard distance between \textsc{x} and \textsc{y}$^\top$ reflects cross-layer reciprocity. } 
  \label{tab:ChemG_stats} 
\end{table}

\paragraph{Results.}
Table \ref{tab:ChemG_stats} shows that the political information network and the perception of influence network are more dense than the scientific information network, while the proportion of ties that are reciprocated is higher in the latter. Notably, there is a large overlap between the political information network and the perception of influence network, meaning that organizations send political information to the organizations they deem influential. Approximately 40\% of the unique ties between these networks overlap, as indicated by a Jaccard index of 0.398. Moreover, organizations tend to send information to those organizations who perceive them to be influential (Jaccard index: 0.308). However, we do not observe such strong association between the scientific information network and the perception of influence network.

We estimated two multiplex \p models on the policy network. Model 1 is a baseline model, while Model 2 includes the effect of governmental status on actors' tendencies to send and receive ties and that of policy preference similarity on within-network density. For both models, we ran four Hamiltonian MCMC chains with 2000 iterations each, employing the priors specified in Section \ref{sec:estimation}. The potential scale reduction statistics $\hat{R}$ are smaller than 1.05 for all the parameters in both models, suggesting model convergence \citep{gelman1992inference}. Further convergence diagnostics for Model 2 can be found in Figures 3--6 in the Supplementary Material \citep{hong2023supplement}. 

Table \ref{tab:chemP_fixed} summarizes the results of the two models. In line with the descriptive statistics, we find a very negative density effect for the scientific information network. Notably, information ties are likely to be reciprocated within and across network layers (log odds: 1.973 for within-layer political, 1.702 for within-layer scientific, 2.492 for cross-layer reciprocity). These findings support Hypothesis 1, suggesting that if actor $i$ delivers any type of information to actor $j$, actor $j$ is more likely to reciprocate with at least one type of information to actor $i$. Furthermore, we observe a positive cross-network density effect between the perception of influence and the political network (log odds: 1.140). This aligns with Hypothesis 2, indicating that actors are more likely to deliver political information to institutions they perceive as influential. In Model 2, we observe a slightly positive effect of policy preference similarity on political network density (log odds: 0.358), suggesting that organizations are more likely to send political information to others with whom they tend to agree on the core policy topics relevant to the chemical regulation process. Additionally, we observe a positive effect of governmental status on receiving political information (log odds: 2.027), indicating that governments receive more political information compared to other types of organizations.

\begin{figure}
    \centering
    \includegraphics[scale=0.45]{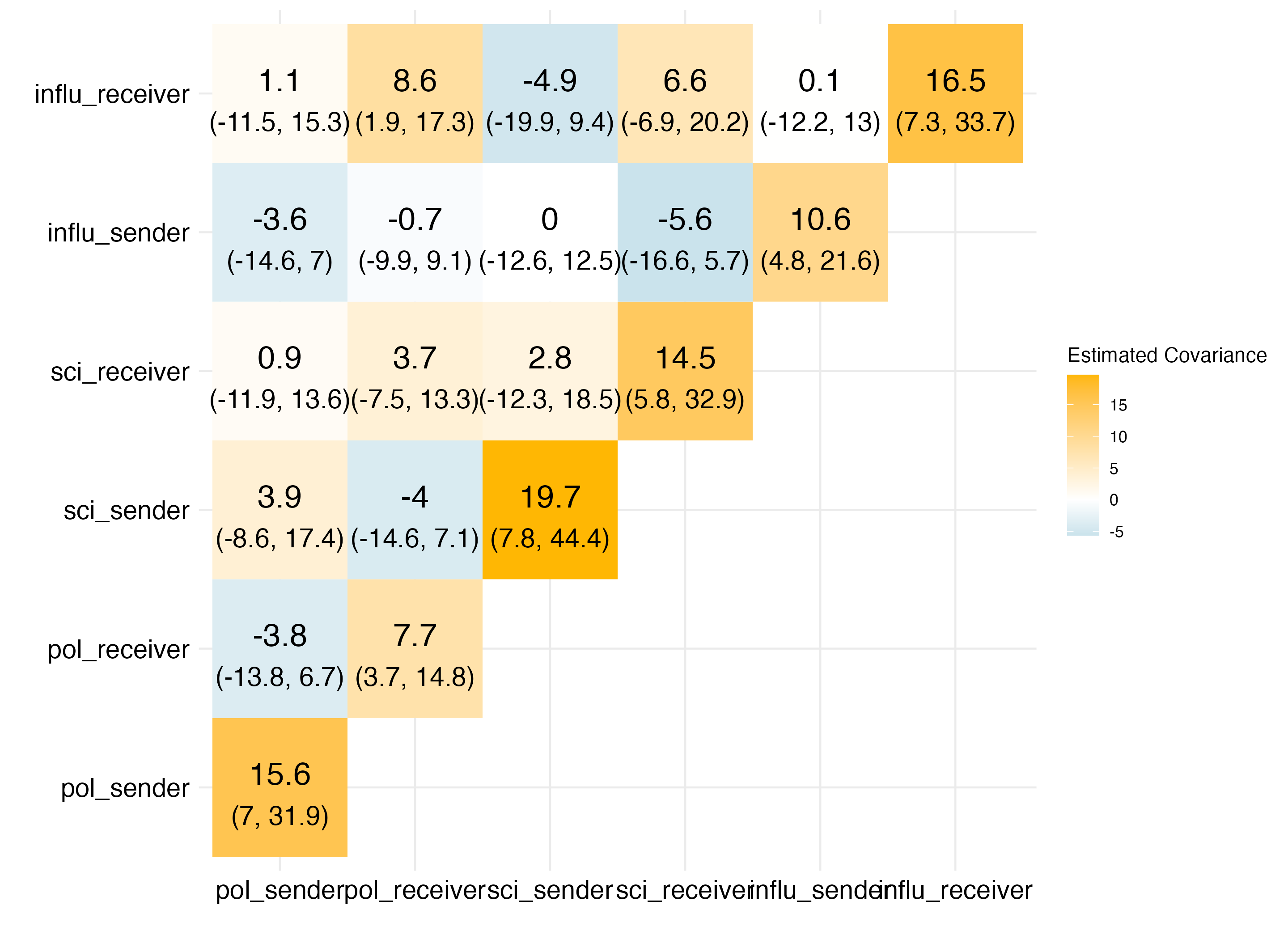}
    \caption{The variance-covariance matrix for the actor random effects in the chemical policy networks (\text{pol} = political information, \text{sci} = scientific information, \text{influ} = perception of influence).}
    \label{fig:Sigma}
\end{figure}

Figure \ref{fig:Sigma} displays the estimated covariance matrix, including the 95\% credible intervals. The diagonal elements represent the variances of the actor effects. The variability in the tendency of organizations to send scientific information ties and receive influence ties is highest. Moreover, the tendency to receive political information ties is positively correlated with the tendency to be perceived as influential (posterior correlation: 0.763). 

\begin{table}[!htbp] \centering 
\begin{tabular}{lrcrc}
\toprule 
& \multicolumn{2}{c}{Model 1} & \multicolumn{2}{c}{Model 2} \\
 & mean\,\, & 95\% CI & mean\,\, & 95\% CI \\
\midrule
\textit{Within-layer dyadic effects} \\
  \,\,\, Density $\mu^\textsc{po}$ & $-$\textbf{2.13} &($-$3.63, $-$0.57) & $-$\textbf{3.80} &($-$5.75, $-$1.76) \\ 
\,\,\,\,\, Preference similarity $\delta_\mu^\textsc{po}$  & &  & \textbf{0.36} &(0.16, 0.57) \\
  \,\,\, Density $\mu^\textsc{sc}$ & $-$\textbf{8.72} &($-$11.43, $-$6.26) & $-$\textbf{9.21} &($-$12.92, $-$5.90) \\ 
  \,\,\,\,\, Preference similarity $\delta_\mu^\textsc{sc}$  & &  & $-$0.11 &($-$0.44, 0.23) \\ 
  \,\,\, Density $\mu^\textsc{pe}$ & $-$1.96 & ($-$4.07, 0.24) & $-$\textbf{3.24} & ($-$5.58, $-$0.99) \\
  \,\,\,\,\, Preference similarity $\delta_\mu^\textsc{pe}$ & &  & 0.07 &($-$0.15, 0.29) \\ 
  \,\,\, Reciprocity $\rho^\textsc{po}$ & \textbf{1.97} &(1.11, 2.84) & \textbf {1.93} &(1.07, 2.80) \\ 
  \,\,\, Reciprocity $\rho^\textsc{sc}$ & \textbf{1.70} &(0.26, 3.24) & \textbf{1.71} &(0.32, 3.22) \\ 
  \,\,\, Reciprocity $\rho^\textsc{pe}$ & 0.14 &($-$0.67, 0.95) & 0.17 & ($-$0.63, 1.00) \\ 
\textit{Cross-layer dyadic effects} \\
 \,\,\, Cross-density $\mu^\textsc{po,sc}$ & \textbf{2.46} & (1.11, 4.02) & \textbf{2.51} & (1.11, 4.12) \\ 
 \,\,\, Cross-density $\mu^\textsc{po,pe}$ & \textbf{1.14} & (0.58, 1.73) & \textbf{1.09} &(0.52, 1.65) \\ 
 \,\,\, Cross-density $\mu^\textsc{sc,pe}$ & 0.31 &($-$0.58, 1.20) & 0.34 &($-$0.55, 1.202) \\ 
 \,\,\, Cross-reciprocity $\rho^\textsc{po,sc}$ & \textbf{2.49} & (1.47, 3.60) & \textbf{2.52} &(1.46, 3.60) \\ 
 \,\,\, Cross-reciprocity $\rho^\textsc{po,pe}$ & $-$0.07 &($-$0.66, 0.52) & $-$0.049 & ($-$0.65, 0.54) \\ 
 \,\,\, Cross-reciprocity $\rho^\textsc{sc,pe}$ & $-$0.17 &($-$1.05, 0.67) & $-$0.17 & ($-$1.05, 0.68) \\ 
  \textit{Sender effects} \\
  \,\,\,Gov. status $\gamma_{\alpha}^\textsc{po}$ & &  & 0.09 & ($-$2.91, 3.02) \\ 
  \,\,\,Gov. status $\gamma_{\alpha}^\textsc{sc}$ & &  & $-$0.75 & ($-$4.20, 2.73) \\ 
  \,\,\,Gov. status $\gamma_{\alpha}^\textsc{pe}$ & &  & 0.65 & ($-$1.79, 3.12) \\ 
  \textit{Receiver effects} \\
  \,\,\,Gov. status $\gamma_{\beta}^\textsc{po}$ & &  & \textbf{2.03} & (0.06, 4.13) \\ 
  \,\,\,Gov. status $\gamma_{\beta}^\textsc{sc}$ &  &  & 2.06 & ($-$0.81, 5.26) \\ 
  \,\,\,Gov. status $\gamma_{\beta}^\textsc{pe}$ & &  & 2.16 & ($-$0.82, 5.22) \\\bottomrule
\end{tabular} 
  \caption{Posterior means and 95\% credible intervals (CIs) for the multiplex \p parameters in the German toxic chemical policy network (\textsc{po} = political information, \textsc{sc} = scientific information, \textsc{PE} = perception of influence). Posterior means with CIs above or below 0 are bolded. Cross-density and cross-reciprocity are short for cross-layer density and cross-layer reciprocity. Model 2 accounts for the effects of preference similarity on network density ($\delta_\mu$) and the actor effects of being a governmental organization ($\gamma_{\alpha}, \gamma_{\beta}$).} 
  \label{tab:chemP_fixed} 
\end{table}

\begin{figure}
    \centering
    \includegraphics[scale=0.5]{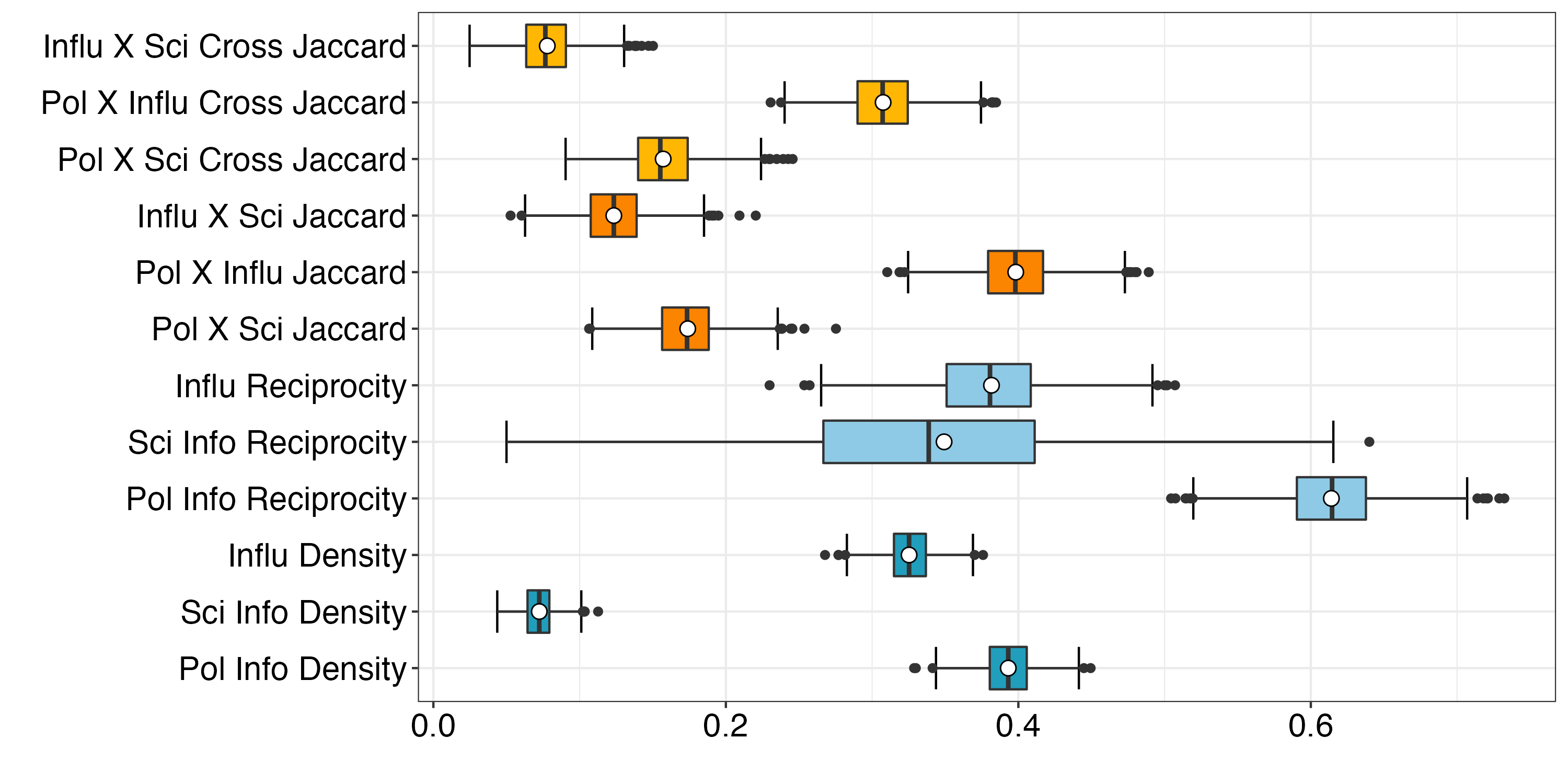}
    \caption{Multiplex goodness-of-fit measures. The white dots indicate the statistics calculated on  the observed network. The box-plots are based on statistics calculated on networks simulated from 1000 posterior draws.}
    \label{fig:multiplex_gof_basic}
\end{figure}

\begin{figure}[ht!]
     \centering
     \begin{subfigure}[b]{\linewidth}
         \centering
         \includegraphics[width=\linewidth]{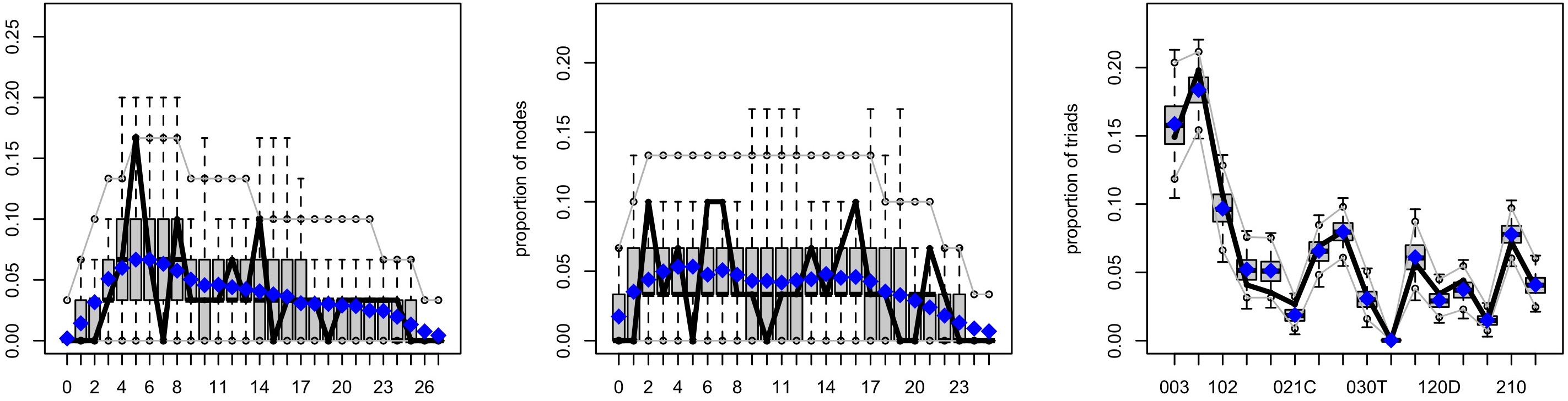}
         \label{fig:pol_gof}
     \end{subfigure}
     \hfill\\
     \begin{subfigure}[b]{\linewidth}
         \centering
         \includegraphics[width=\linewidth]{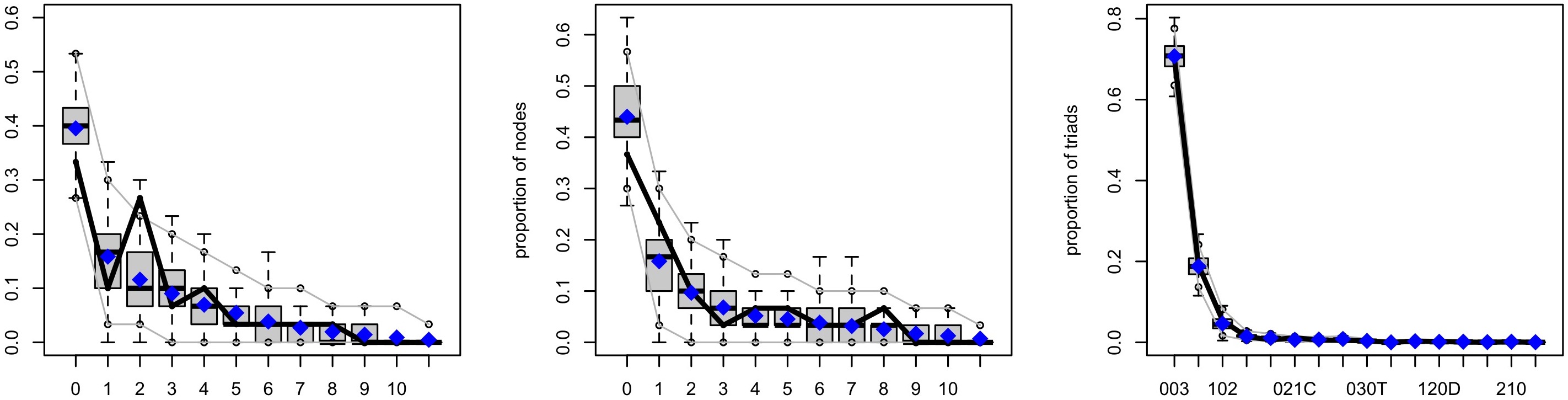}
         \label{fig:sci_gof}
     \end{subfigure}
     \hfill\\
     \begin{subfigure}[b]{\linewidth}
         \centering
         \includegraphics[width=\linewidth]{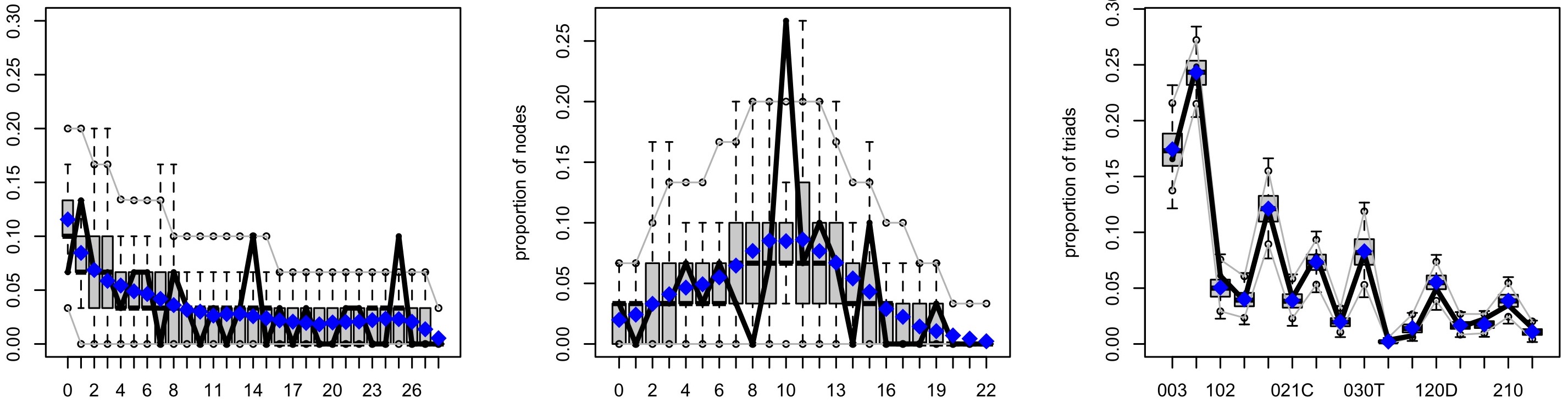}
         \label{fig:infrep_gof}
     \end{subfigure}

    \caption{Simulated goodness-of-fit statistics for the policy triplex network: political information (top row), scientific information (middle), and perception of influence (bottom). From left to right, the statistics are the indegree distribution, the outdegree distribution, and the triad census.
    The figure compares the observed network's statistics (black line) to the distribution of the same statistics on networks generated from the fitted model (blue dots: simulated average).}
    \label{fig:ergm_gof}
\end{figure}

\paragraph{Goodness-of-fit.}
We apply the goodness-of-fit methods described in Section \ref{sec:gof} to Model 2. Figure \ref{fig:ergm_gof} shows the results of the existing uniplex goodness-of-fit methods with implementation adapted from the \texttt{latentnet} package \citep{krivitsky2015package}. Overall, we observe a good fit for the indegree and outdegree distributions, as well as for the triad census, in each layer of the network individually. The adequate fit of the triad census suggests that the \p model is appropriate for these data, despite not explicitly modeling triadic effects. Furthermore, Figure \ref{fig:multiplex_gof_basic} displays the results of the proposed multiplex goodness-of-fit methods. The results demonstrate an adequate fit to the multiplex network. Additional goodness-of-fit checks on the covariance among the in- and outdegrees of actors in the three networks can be found in the Figure 7 in the Supplementary Material \citep{hong2023supplement}.

\newpage
\section{Application: Discrepancies in perceptions of gossip}
\label{sec:gossip}

Gossip is a universal phenomenon in human groups. It can be defined as informal communication about a third, non-present person \citep{dores2021integrative} and has been linked to both positive outcomes, such as promoting cooperation, and negative outcomes, such as the social exclusion of the target \citep[e.g.,][]{feinberg2014gossip, kisfalusi2019adolescent}. In this section, we focus on gossip in the school setting. Here, gossip has been considered a form of bullying \citep{kisfalusi2018bullies}, which significantly affects children's social development and school outcomes.

Despite the important role of gossip in social interactions in schools, there are nuances to the phenomenon which are often overlooked. In particular, studies on gossip usually only consider the viewpoint of either the gossiper or the gossip target. Yet, different individuals may perceive the same event differently, leading to divergent perspectives. In the case of bullying, for example, individuals who are identified as bullies by their victims may not self-identify as such. The non-confrontational nature of gossip introduces an additional layer of complexity when gossip targets need to identify who they think is talking about them behind their back. Previous research has examined the disagreements between bullies and victims in reported bullying behavior \citep{Veenstra2007, Tolsma2013, kisfalusi2018bullies, Tatum2020}. In this setting, the two perspectives were highly complementary. Yet, the disagreements between gossipers and gossip targets so far remain unstudied.

\begin{figure}[b!]
     \centering
     \begin{subfigure}[b]{0.4\linewidth}
         \centering
         \includegraphics[width=\linewidth]{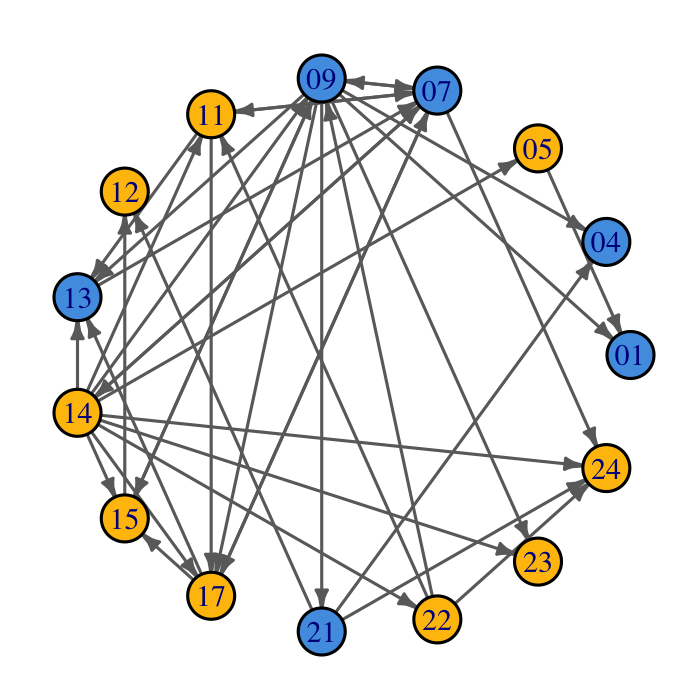}
         \caption{\textbf{Gossip network}: $i \to j$ means $i$ admits to gossip about $j$.}
         \label{fig:gossip}
     \end{subfigure}
     \hfill
     \begin{subfigure}[b]{0.4\linewidth}
         \centering
         \includegraphics[width=\linewidth]{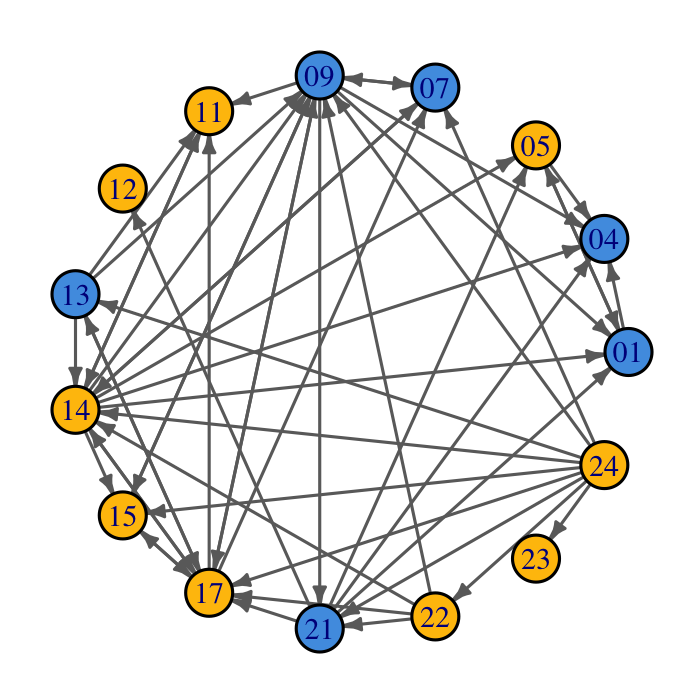}
         \caption{\textbf{Perception network}: $i \to j$ means $i$ thinks $j$ gossips about them.}
         \label{fig:perception}
     \end{subfigure}
     \hfill\\
     \begin{subfigure}[b]{0.4\linewidth}
         \centering
         \includegraphics[width=\linewidth]{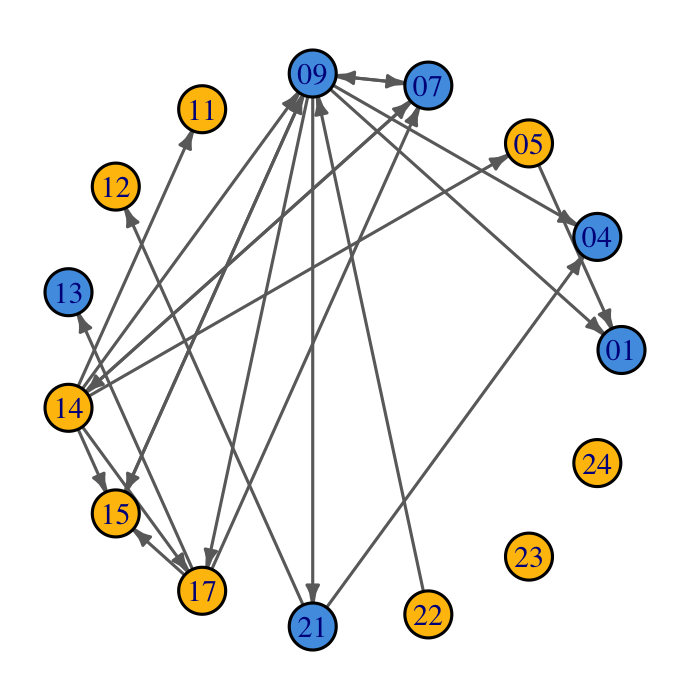}
         \caption{\textbf{Retaliation}: $i \to j$ means $i$ thinks $j$ gossips about them \emph{and} $i$ admits to gossip about $j$.}
         \label{fig:retaliation}
     \end{subfigure}
     \hfill
     \begin{subfigure}[b]{0.4\linewidth}
         \centering
         \includegraphics[width=\linewidth]{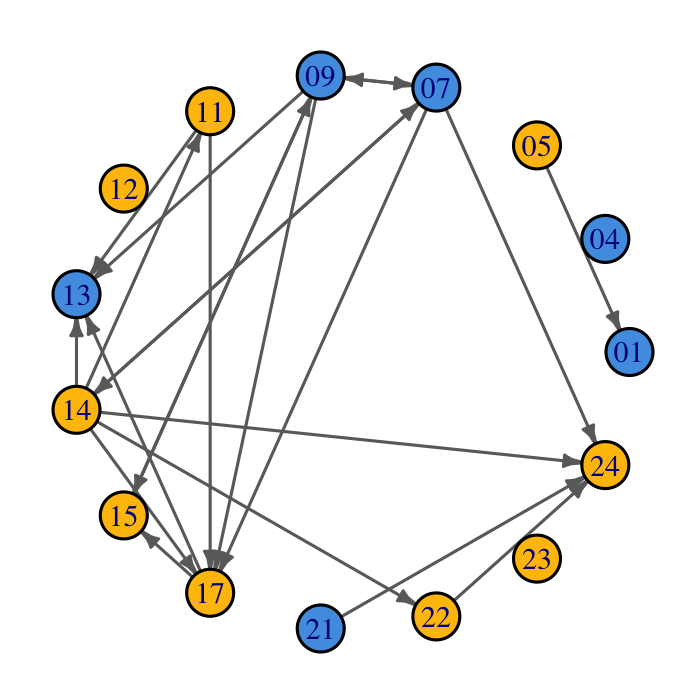}
         \caption{\textbf{Accurately perceived gossip}: $i \to j$ means $i$ admits to gossip about $j$ \emph{and} $j$ accurately perceives $i$ as gossiping about them.}
         \label{fig:accurate_perception}
     \end{subfigure}
    \caption{An example of a biplex network of self-reported gossip ties and perceived gossip ties among 15 students. There are 9 female (yellow) and 6 male students (blue).}
    \label{fig:gossip_network_vis}
\end{figure}

In this section, we take on a multiplex network perspective on the gossip relation, simultaneously studying self-reported gossip behavior and the perceptions of gossip behavior by the gossip target.  Being a gossiper might affect how an individual perceives themselves as the victim of gossip, and gossip victims may feel the need to retaliate.  A multiplex network perspective is necessary to understand such associations. In particular, we will focus on the following two research questions. 
First, are students more likely to gossip about people who they think are gossiping about them? We expect this is the case and will refer to this as \emph{retaliation}. Second, are students likely to accurately perceive their gossipers? As discrepancies between reports by aggressors and victims have been found in the case of bullying \citep{Veenstra2007}, we expect these to exist as well in reports of gossip. We will refer to this as \emph{perception discrepancy} (as opposed to \emph{perception accuracy}). In the case of retaliation, individual $i$ both gossips about $j$ \emph{and} thinks that $j$ is gossiping about them. Accurately perceived gossip ties are those where individual $i$ gossips about $j$, and $j$ thinks that $i$ gossips about them. Figure \ref{fig:gossip_network_vis} illustrates these relations.

We expect that gender moderates the above effects, such that the likelihood of retaliation and accurate perception is higher when the sender and the receiver are of the same gender. To investigate this hypothesis, we incorporate the binary indicator of whether the two actors in a dyad share the same gender as a covariate when examining cross-network density and reciprocity. Additionally, we study how gender influences individuals' propensities to both send and receive ties by incorporating gender as an actor covariate in our analysis.

\paragraph{Data.} We will study the gossiper-target relationship based on data from the fourth wave of a six-wave panel study on Hungarian elementary school students, conducted between 2013 and 2017. The fourth wave of data collection occurred in the spring of 2015, when the students were enrolled in the sixth grade and were on average 13 years old. We dropped 9 classes with zero perception ties and analyze the remaining 34 classes containing a total of 702 students. See \cite{kisfalusi2018bullies, Kisfalusi2021Grading} for more detailed information on the data collection. We simultaneously study self-reported gossip (students' answers to the question: `About whom do you talk with your classmates behind his/her back?') and perceived gossip (`Who do you think is talking about you with other classmates behind your back?'). These two relations together yield a biplex network. Figure \ref{fig:gossip_network_vis} illustrates the gossip data collected in one of the classrooms. An example of a retaliation-only dyad is $\{9, 21\}$, where student 9 believes that student 21 is gossiping about him, and retaliates by gossiping about 21. However, 21 neither gossips about 9 nor accurately perceives that 9 is gossiping about him. 

Descriptive statistics of the 34 biplex networks are summarized in Table \ref{tab:gossip_stats}. On average, we have about 20 students present in each classroom with a balanced proportion of female to male students. On average, the gossip networks have fewer ties than the perception networks, and reciprocity is also lower in the gossip networks.  Based on the Jaccard indices, there are considerably more pairs of students $i$ and $j$ where $i$ gossips about $j$ and $i$ thinks $j$ is gossiping about them than pairs where $i$ gossips about $j$ and $j$ accurately perceives that $i$ is gossiping about them. This descriptive finding concurs with our expectation on students' retaliating behavior. 
\begin{table}[htb]
\centering
\begingroup\fontsize{9pt}{10pt}\selectfont
\begin{tabular}{lrrr}
  \hline
  & Mean & Median & SD \\
 \hline
   Number of students & 20.7 & 19.5 & 4.96 \\ 
  Proportion female & 0.48 & 0.47 & 0.10 \\ 
Gossip density & 0.09 & 0.08 & 0.06 \\ 
  Gossip reciprocity & 0.13 & 0.14 & 0.11 \\ 
  Perception density & 0.15 & 0.14 & 0.06 \\ 
  Perception reciprocity & 0.23 & 0.23 & 0.11 \\ 
  Gossip $\times$ perception Jaccard & 0.24 & 0.21 & 0.12 \\ 
  Gossip $\times$ perception$^\top$ Jaccard & 0.11 & 0.10 & 0.07 \\
  \hline
\end{tabular}
\endgroup
\caption{Summary statistics of the gossip and perception of gossip networks in 34 Hungarian elementary school classrooms.} 
\label{tab:gossip_stats}
\end{table}

\vspace{-4mm}
\paragraph{Plan of Analysis.}
We fit two models to each classroom: a baseline biplex \p model and a model that additionally includes the effect of gender (female\,=\,1) on students' tendencies to send and receive ties, and the (dyadic) effect of having the same gender on cross-layer density and reciprocity. For each model, we used MCMC sampling with 4 chains with 1600 iterations, including 800 warmup iterations. Posterior estimates are calculated based on 3200 posterior draws (800 post-warmup draws from each of the 4 chains). The potential scale reduction statistic $\hat{R}$ was less than 1.05 for all the parameters in each of the 68 models, indicating the convergence of the models. 

To aggregate the results from all classrooms, we conducted a Bayesian meta-analysis by fitting a normal-normal hierarchical model to the parameter estimates of each classroom \cite[Ch.\,5.5]{gelman2013bayesian}. Let $\bm{\hat{\theta}} \in \real{34}$ denote the vector of the posterior means for one of the model parameters. We assume that
\begin{equation*}
    \begin{split}
        \hat{\theta}_j &\sim \mathcal{N}(\theta_j, \sigma^2_j), \\
        \theta_j &\sim  \mathcal{N}(\mu, \tau^2),
    \end{split}
\end{equation*}
where $\mu$ and $\tau$ are the overall population mean and standard deviation. For the priors of the hyperparameters, we assume $\mu \sim \mathcal{N}(0, 10)$ and $\tau \sim \text{Cauchy}(0, 0.5)$ for all the fixed effects. For the covariance of the random sender and receiver effects, we choose $\mu \sim \mathcal{N}(0, 100)$ and $\tau \sim \text{Cauchy}(0, 0.5)$. We used the \texttt{R} package \texttt{brms} \citep{brms2021Bürkner} with 5000 iterations for the meta-analysis \citep[Ch.\,13]{harrer2021doing}.

\paragraph{Results.} Table \ref{tab:meta_res} shows the posterior means and the 95\% credible intervals of the population means of the \p model parameters. In Model 1, we find a positive cross-density effect (estimated population mean $\hat\mu = 2.62$), which supports our expectation about gossip retaliation. Students are about 13 times more likely to gossip about someone if they think the person is gossiping about them (or vice versa). We do not observe a positive cross-reciprocity effect $(\hat\mu=-0.13$) indicating no evidence that students accurately perceive who is gossiping about them. This result is consistent with our expectation on gossip perception and the studies on the dual perspective on bullying. 

Gossip networks are generally sparse, and we indeed find negative network-specific density effects in Model 1 for both  self-reported and perceived gossip. Negative ties are often reciprocated but, interestingly, we do not observe a reciprocity effect for self-reported gossip ($\hat\mu = -0.72$; 95\% credible interval $[-2.11,0.67]$). Yet, for perceived gossip, we do find evidence of positive reciprocity. A reason for this may be that students feel more comfortable reporting on perceived negative behavior than on their own negative behavior, because of social desirability bias. Also, if two students dislike each other, they may each suspect the other to be gossiping about them. As such, a reciprocated perceived gossip relation could be indicative of a generally negative tie.

After accounting for effects of gender in Model 2, the above-mentioned findings remain the same. 
We find a positive same-gender cross-density effect ($\hat\mu = 1.99$), indicating that $i$ is about 7 times more likely to gossip about $j$ \emph{and} to think $j$ is gossiping about them if $i$ and $j$ are of the same gender. We do not find evidence that perception accuracy changes when gossiper and target are of the same gender. Also, controlling for the other effects, we do not find a differential tendency for female students to send or receive self-reported or perceived gossip ties.  

Finally, we consider the estimated covariance of the actor random effects, for which the two models yield comparable results. In particular, the gossip network sender random effect is positively correlated with the perception network sender random effect and the same holds true for the receiver random effects in both networks. This means that individuals who report to gossip a lot about others also tend to think others are gossiping about them -- a tendency that goes beyond the urge to retaliate among specific student pairs. And individuals who are gossiped about a lot are also suspected to be avid gossipers. Note that the first finding is a negative form of generalized reciprocity, where individuals treat others in the same way that others treated them in the past. 
Interestingly, we find a negative covariance between the perception network random sender and receiver effects. That is, the more students think they are gossiped about, the less likely they are accused of gossip by others. This finding contrasts the positive perception reciprocity effect we found at the dyad level. 

\begin{table}[!htbp] \centering 
\begin{tabular}{lrcrc}
\toprule 
& \multicolumn{2}{c}{Model 1} & \multicolumn{2}{c}{Model 2} \\
 & mean\,\, & 95\% CI & mean\,\, & 95\% CI \\
\midrule
\textit{Within-layer dyadic effects} \\
\,\,\, Density $\mu^\textsc{g}$ & $-$\textbf{10.05} &($-$11.48, $-$8.52) & $-$\textbf{11.02} &($-$12.32, $-$9.39) \\  
 \,\,\, Density $\mu^\textsc{p}$ & $-$\textbf{5.99} &($-$6.67, $-$5.35) & $-$\textbf{7.21} &($-$8.25, $-$6.22) \\ 
 \,\,\, Reciprocity $\rho^\textsc{g}$ & $-$0.72 &($-$2.11, 0.67) & $-$0.91 &($-$2.46, 0.65) \\
\,\,\, Reciprocity $\rho^\textsc{p}$  & \textbf{1.38} &(0.65, 2.07) & \textbf{1.472} &(0.69, 2.20) \\ 
\textit{Cross-layer dyadic effects} \\
 \,\,\, Cross-density $\mu^\textsc{g,p}$ &\textbf{2.62} &(1.78, 3.48) & \textbf{1.47} &(0.47, 2.46) \\ 
\,\,\,\,\, Same-gender $\delta^\textsc{g,p}_{\mu, \text{cross}}$ &  & &\textbf{1.99} &(0.66, 3.22) \\ 
 \,\,\, Cross-reciprocity $\rho^\textsc{g,p}$ & $-$0.13 &($-$0.98, 0.79) & $-$0.93 &($-$2.15, 0.19) \\ 
\,\,\,\,\, Same-gender $\delta^\textsc{g,p}_{\rho, \text{cross}}$ &  & &$-$0.45 &($-$2.04, 1.14) \\ 
  \textit{Sender effects} \\
\,\,\, Female $\gamma_{\alpha}^\textsc{g}$ &  & & $-$0.11 &($-$1.90, 1.71) \\ 
\,\,\, Female $\gamma_{\alpha}^\textsc{p}$ & &  & 0.66 &($-$0.46, 1.79) \\ 
  \textit{Receiver effects} \\
\,\,\, Female $\gamma_{\beta}^\textsc{g}$ & &  & $-$0.66 &($-$1.74, 0.37) \\ 
 \,\,\, Female $\gamma_{\beta}^\textsc{p}$ &  &  & $-$0.45 &($-$2.04, 1.14) \\ 
\textit{Covariance} \\
\,\,\, Sender$^\textsc{g}$$\times$\,receiver$^\textsc{g}$ $\sigma_{A^\textsc{g},B^\textsc{g}}$  & $-$0.57 &($-$4.25, 2.98) & 0.20 &($-$4.50, 5.25) \\ 
\,\,\, Sender$^\textsc{g}$$\times$\,sender$^\textsc{p}$ $\sigma_{A^\textsc{g},A^\textsc{p}}$ & \textbf{17.33} &(7.63, 26.77) & \textbf{24.13} &(4.71, 42.75) \\ 
\,\,\, Sender$^\textsc{g}$$\times$\,receiver$^\textsc{p}$ $\sigma_{A^\textsc{g},B^\textsc{p}}$  & 1.12 &($-$2.77, 5.12) & $-$1.56 &($-$5.18, 1.81) \\ 
\,\,\, Sender$^\textsc{p}$$\times$\,receiver$^\textsc{g}$ $\sigma_{A^\textsc{p},B^\textsc{g}}$ & 2.16 &($-$1.23, 5.51) & 3.49 &($-$0.84, 7.71) \\
\,\,\, Receiver$^\textsc{g}$$\times$\,receiver$^\textsc{p}$ $\sigma_{B^\textsc{g},B^\textsc{p}}$ & \textbf{4.59} &(1.82, 7.13) & \textbf{3.77} &(1.25, 6.22) \\ 
\,\,\, Sender$^\textsc{p}$$\times$\,receiver$^\textsc{p}$ $\sigma_{A^\textsc{g},B^\textsc{p}}$  & $-$\textbf{3.54} &($-$7.06, $-$0.33) & $-$\textbf{5.08} &($-$8.58, $-$1.78) \\
  \bottomrule
\end{tabular} 
  \caption{Posterior means and 95\% credible intervals (CIs) of the overall means of the multiplex gossip networks \p parameters from 34 Hungarian elementary school classrooms (\textsc{g} = gossip, \textsc{p} = perception of gossip). Posterior means with CIs above or below 0 are bolded. Cross-density and cross-reciprocity are short for cross-layer density and cross-layer reciprocity. Model 2 accounts for the effects of gender on cross-layer density ($\delta^\textsc{g,p}_{\mu, \text{cross}}$) and reciprocity ($\delta^\textsc{g,p}_{\rho, \text{cross}}$) and the actor effects of being female ($\gamma_{\alpha}, \gamma_{\beta}$).} 
  \label{tab:meta_res}
\end{table}

\section{Discussion}
\label{sec:disc}
As social actors are often embedded in multiple interconnected social networks, we propose a Bayesian multiplex network model in the \p modeling framework. This model captures the interplay of social dynamics across different network layers by introducing cross-layer dyadic effects and actor random effects. By assuming the dyads are conditional independent, we can formulate the model as a mixed-effects multinomial logistic regression capable of handling a range of network dependencies while remaining interpretable. The proposed methodology is available in the R package \texttt{multip2}. This package also includes the data to replicate the analyses presented in Section \ref{sec:pol}.

Despite being widely used and simplifying model specification and estimation, the conditional independent dyads assumption also is a limitation of the \p model. This assumption implies that the probability of a tie between actors $i$ and $k$ is unaffected by the presence of a tie between actors $i$ and $j$ and between $j$ and $k$, even though in real networks, transitivity (e.g., befriending the friend of your friend) is quite common. Our definition of the multiplex \p model does not explicitly represent triadic effects. Nonetheless, as illustrated in Figure \ref{fig:ergm_gof}, the model adequately captures the triadic structures present in the policy network. Furthermore, as shown in Figure 8 of the Supplementary Material, the  multiplex \p model captures multiplex goodness-of-fit statistics, such the overlap between different networks, much better than a uniplex approach, where a \p model is fitted on each of the network layers separately. 

The ability of lower-order network models to capture higher-order effects (i.e., triad effects) has been documented and explored by \cite{faust2010puzzle}. 
In the context of the social relations model, \cite{minhas2019inferential} accounted for triadic effects while preserving the assumption of conditional independent dyads by incorporating a multiplicative component in the form of a latent factor model in their model. The multiplex \p model could be extended similarly. Moreover, while we have proposed several multiplex goodness-of-fit measures, it still remains unexplored how to best incorporate goodness-of-fit statistics that capture multiplex triadic patterns. As the number of network layers increases, the number of possible triad configurations on a dyad grows exponentially. Therefore, the choice of which multiplex network triad configurations to include should be guided by the specific applied research context. 

Additionally, it is important to note that the priors employed in our study remain invariant to increasing network dimensions and network size across key network statistics, such as density, reciprocity, and transitivity. We have included additional prior predictive checks in Figure 2 of the Supplementary Material \citep{hong2023supplement}. However, when applying the multiplex \p model, practitioners should conduct separate prior predictive checks to ensure that the selected priors induce appropriate network properties. 

Finally, in the gossip study (Section \ref{sec:gossip}), we considered data from 34 of school classes and aggregated the multiplex \p model results by a Bayesian meta-analysis. Nowadays, network data collection frequently involves sampling from a population of networks \citep[e.g., school classes, households, organizational work units;][]{goeyvaerts2018household, lubbers2003}. In such samples, bigger networks typically have lower density, while the mean degree is similar across the networks \citep{krivitsky2011adjusting} -- something we did not account for in the current study. A multilevel extension of the multiplex \p model would enable researchers to study a sample of multiplex networks across various groups, and thus obtain more generalizable results and address group-level research questions. Moreover, the multilevel approach would estimate parameters more efficiently by partially pooling observations across the groups to estimate the global parameters. The multilevel model could also explicitly account for the effect of network size variation in a sample of networks on parameter estimates, in case that variation was large \citep{krivitsky2023tale, niezink2023discussion}.

\newpage
\begin{acks}[Acknowledgments]
The authors would like to thank K\'aroly Tak\'acs for sharing the data analyzed in Section \ref{sec:gossip} and for the helpful discussions. 
\end{acks}

\begin{funding}
This work was supported by NSF Grant No. SES-2020276. 
\end{funding}

\begin{supplement}
\stitle{Supplement to "The Multiplex \p Model: Mixed-Effects Modeling for Multiplex Social Networks"}
\sdescription{The Supplementary Material contains additional prior predictive checks, convergence checks, and goodness-of-fit results.}
\end{supplement}

\bibliographystyle{ba}
\bibliography{references}




\section*{Supplementary material}

# Supplement to "The Multiplex $p_2$ Model: Mixed-Effects Modeling for Multiplex Social Networks"

Anni Hong*, Nynke M.D. Niezink*

**Additional prior predictive checks.** We conducted additional prior predictive checks comparing the inverse Wishart prior to the LJK prior on the variance-covariance matrix for random actor effects. As shown in Figure 1, draws from the inverse Wishart prior yield resonable uniplex networks but extreme biplex networks. On the other hand, Figure 2 shows that LJK prior yields a reasonable distribution for the statistics for a variety of multiplex networks (i.e., a biplex network with 100 nodes and a 5-layer network with 30 nodes).

**Convergence checks for Application 1.** We present evidence of model convergence in the study of information exchange in a policy network by examining the traceplots and the Geweke $z$-scores. We can monitor the convergence and the mixing of the MCMC chains visually by assessing the plotted samples per chain at each iteration, i.e., the traceplot. When the model has converged, the samples will exhibit no trends, displaying horizontal bands as shown in Figure 3 and Figure 4.

In Figure 5 and Figure 6, we display the the Geweke diagnostic of model convergence, which assesses $z$-scores derived from the means of the chains' initial and final segments (by default, the first 10% and the last 50%). Convergence is indicated by small absolute $z$-scores, suggesting that the two segments are indistinguishable from each other. Both in Figure 5 and 6, we observe that roughly 95% of $z$-scores in the chains fall within the $-2$ to 2 range, suggesting model convergence.

**Additional goodness-of-fit results for Application 1.** Figure 7 presents multiplex goodness-of-fit checks on the covariance among the in- and outdegrees of actors in the three networks in Application 1. Networks are simulated using the multiplex $p_2$ fit of Model 2. As shown by the figure, the fitted degree covariance largely agrees with the observed covariance.

We demonstrate that the multiplex $p_2$ model can lead to better fit than fitting individual $p_2$ model to each layer of the network separately, by capturing cross-network dependencies. Figure 8 shows that the uniplex approach yields worse fit on cross-layer dyadic goodness-of-fit measures (Figure 8a), and slightly worse fit on the cross-layer degree covariance (Figure 8b).

---

*Department of Statistics and Data Science, Carnegie Mellon University, Pittsburgh, PA 15213
annihong@andrew.cmu.edu nniezink@andrew.cmu.edu

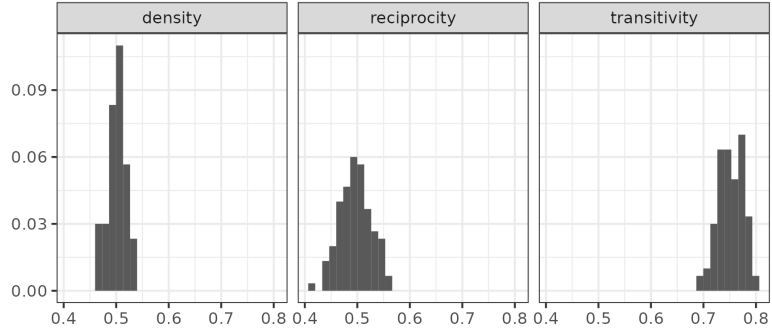

(a) Prior predictive checks calculated based on 100 simulated uniplex networks, where $\mathbf{\Sigma}_{AB} \sim \text{InverseWishart}(I_{2t}, \nu = 3)$.

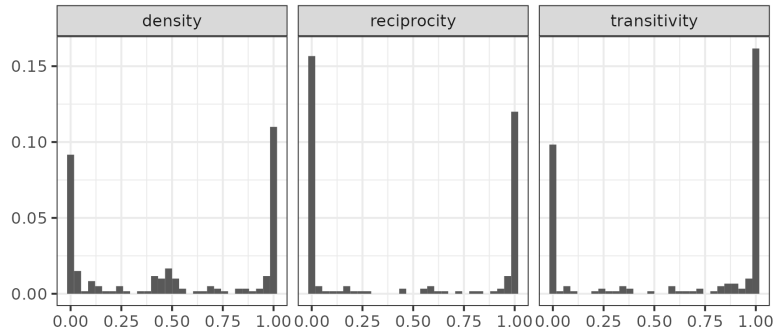

(b) Prior predictive checks calculated based on 100 simulated biplex networks, where $\mathbf{\Sigma}_{AB} \sim \text{InverseWishart}(I_{2t}, \nu = 5)$.

Figure 1: The inverse Wishart prior does not scale with the number of network layers and already yields extreme networks in simulations of a biplex network. Simulations were done on networks of size $n = 30$.

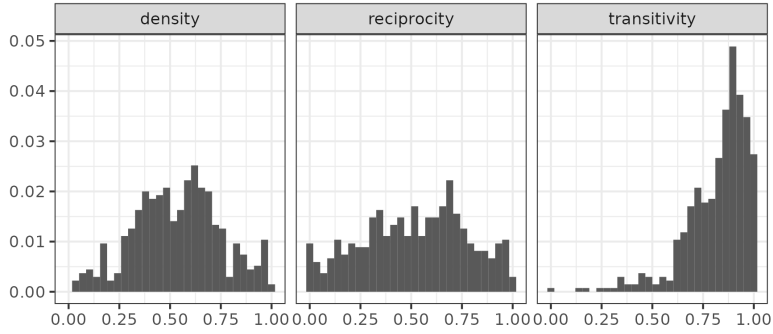

(a) Prior predictive checks calculated based on 100 simulated 5-layer multiplex networks with $n = 30$ nodes.

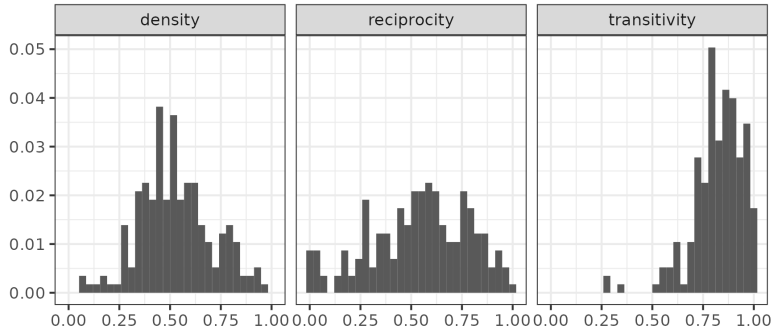

(b) Prior predictive checks calculated based on 100 simulated biplex networks with $n = 100$ nodes.

Figure 2: The prior predictive checks on the networks simulated using LJK priors for the correlation matrix and an inverse gamma prior for the elements of the scale vector do are not very much affected by increasing network the dimension (2 versus 5 layers) or the number of network actors (30 versus 100).

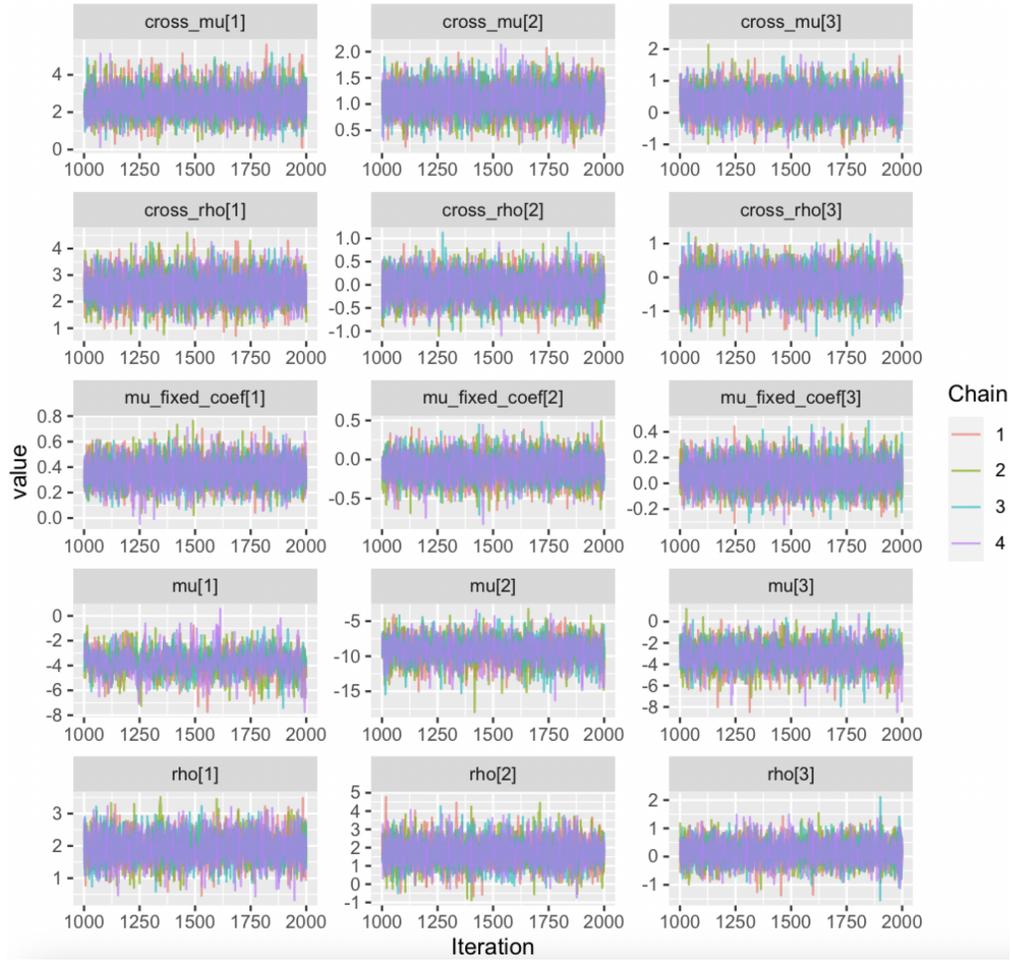

Figure 3: Traceplots of the fixed effect parameters in Application 1.

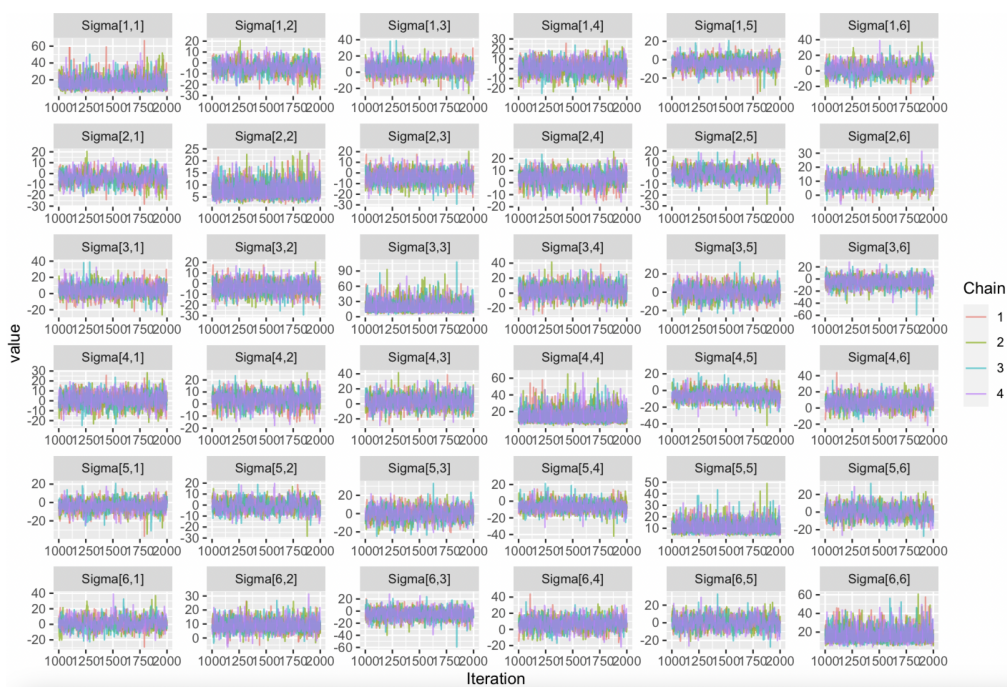

Figure 4: Traceplots of the random effect parameters in Application 1.

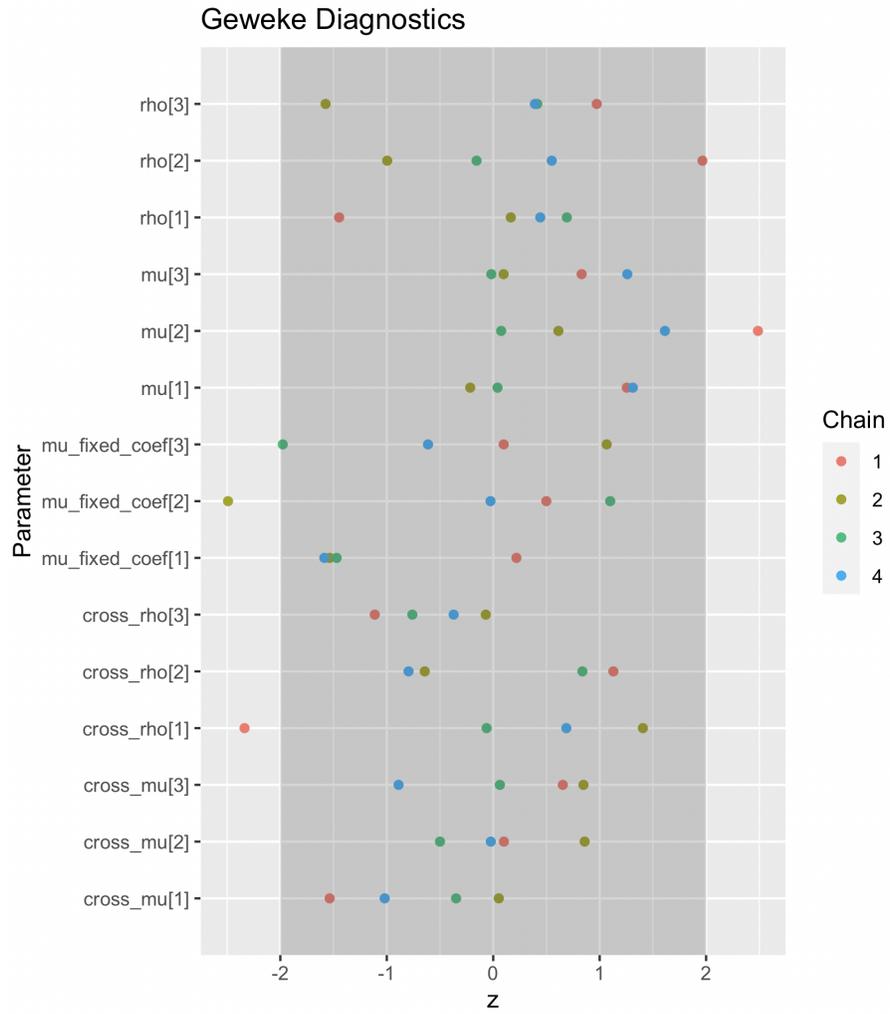

Figure 5: Geweke diagnostic plot for the fixed parameters in Application 1. Most $z$-scores fall within the $-2$ to $2$ range (shaded region), signifying convergence in the chains.

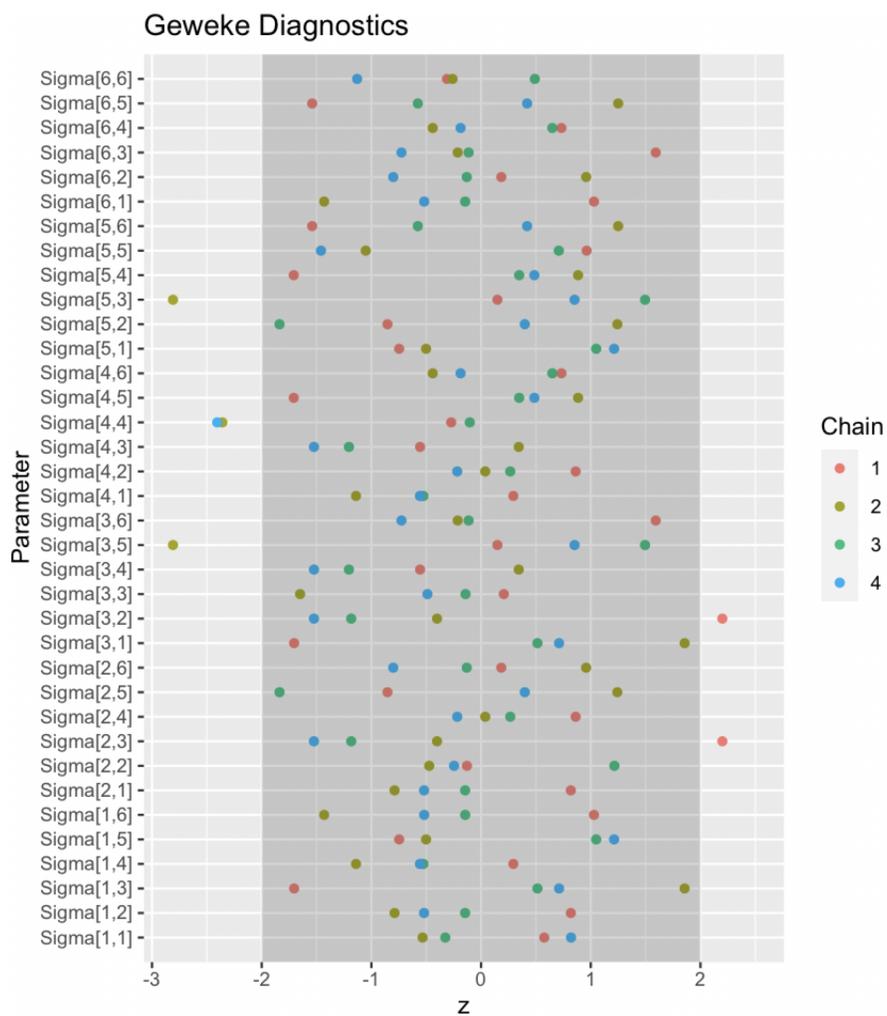

Figure 6: Geweke diagnostic plot for the covariance of the random effects in Application 1. Most $z$-scores fall within the $-2$ to $2$ range (shaded region), signifying convergence in the chains.

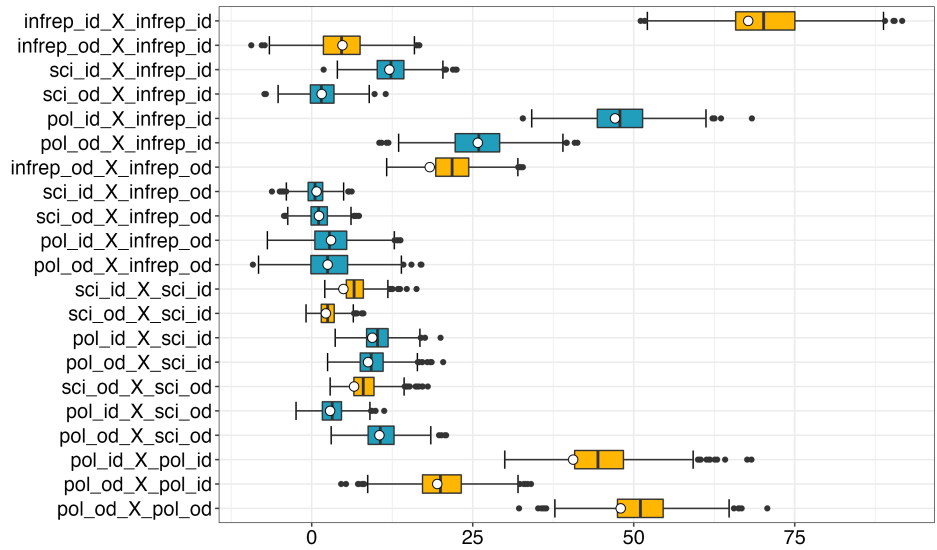

Figure 7: Multiplex goodness-of-fit on the covariance among the in-degree (id) and the out-degree (od) distributions for the three network layers: perception of influence (infrep), scientific information exchange (sci), and political info exchange (pol). The white dots indicate statistics calculated on observed networks. The boxplots are based on statistics calculated on networks simulated from 1000 posterior draws. Within-layer covariance statistics are shown in yellow, and cross-layer covariance statistics are shown in blue.

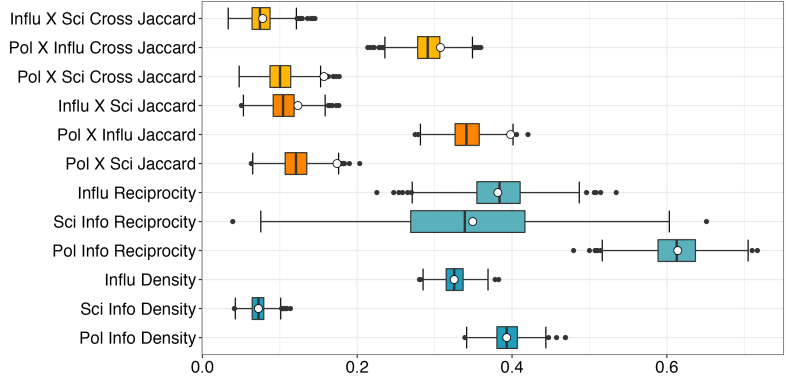

(a) Networks simulated from the uniplex $p_2$ model show significantly worse fit to multiplex goodness-of-fit measures (i.e., the Jaccard indices indicating the overlap between networks).

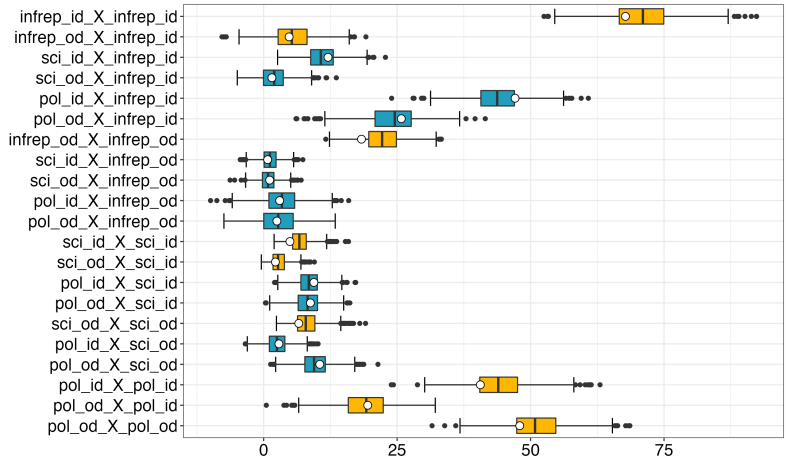

(b) Networks simulated from the uniplex $p_2$ model show slightly worse fit to multiplex goodness-of-fit measures on the cross-layer correlations (shown in blue).

Figure 8: By fitting a $p_2$ model on each layer of the network separately, we obtain significantly worse fit on cross-layer dyadic goodness-of-fit measures (Figure 8a), and a slightly worse fit on the cross-layer degree covariance (Figure 8b).



\end{document}